\documentclass[preprint]{aastex}
\usepackage{natbib,graphicx,latexsym}
\shorttitle{New optical and near-infrared SBF models}
\shortauthors{Cantiello M. et al.}

\newcommand{\lsim}{\ \raise
-2.truept\hbox{\rlap{\hbox{$\sim$}}\raise5.truept\hbox{$<$}\ }}
\newcommand{\gsim}{\ \raise
-2.truept\hbox{\rlap{\hbox{$\sim$}}\raise5.truept\hbox{$>$}\ }}

\begin{document}

\title{New optical and near-infrared Surface Brightness Fluctuations models.
A primary distance indicator ranging from Globular Clusters 
to distant galaxies?}

\author{
Cantiello M.\altaffilmark{1,2},
Raimondo G.\altaffilmark{1,3},
Brocato E.\altaffilmark{1,4},
Capaccioli M.\altaffilmark{5,6}
}

\altaffiltext{1}{INAF--Osservatorio Astronomico di Collurania, Via
M. Maggini, I-64100 Teramo, Italy}

\altaffiltext{2} {Dipartimento di Fisica ``E.R. Caianiello'',
Universit\`a di Salerno and INFN, Sezione di Napoli, Gruppo Collegato
di Salerno, Via S. Allende, 84081 Baronissi, Salerno, Italy; cantiello@te.astro.it}

\altaffiltext{3}{Dipartimento di Fisica, Universit\`a La Sapienza,
P.le Aldo Moro 2, I-00185 Roma, Italy; raimondo@te.astro.it}

\altaffiltext{4}{Istituto Nazionale di Fisica Nucleare, LNGS,
L'Aquila, Italy; brocato@te.astro.it}

\altaffiltext{5}{Dipartimento di Scienze Fisiche, Universit\`a di Napoli
Federico II, Complesso Monte S. Angelo, via Cintia, I-80126,
Napoli, Italy; capaccioli@na.astro.it}

\altaffiltext{6}{INAF--Osservatorio Astronomico di Capodimonte, via
Moiariello 16, I-80131 Napoli, Italy}

\begin{abstract}

We present new theoretical models for Surface Brightness
Fluctuations (SBF) both for optical and near-infrared bands in
standard ground-based and Hubble Space Telescope filter
systems. Simple Stellar Population simulations are adopted. Models
cover the age and metallicity ranges from $t=5$ to $15~Gyr$ and from
$Z=0.0001$ to $0.04$ respectively. Effects due to the variation of the
Initial Mass Function and the stellar color-temperature relations are
explored. Particular attention is devoted to very bright stars in the
color-magnitude diagram and to investigate the effects of mass loss
along the Red Giant Branch (RGB) and the Asymptotic Giant Branch
(AGB). It is found that $U$ and $B$ bands SBF amplitudes are powerful
diagnostics for the morphology of the Horizontal Branch and the
Post-AGB stars population. We point out that a careful treatment of
mass loss process along the RGB and AGB is fundamental in determining
reliable SBF evaluations.

The SBF measurements are used to give robust constraints on the evolution
of AGB stars, suggesting that mass loss activity on AGB
stars should be twice more efficient than on the RGB stars.

Our models are able to reproduce the absolute SBF magnitudes of the
Galactic Globular Clusters and of galaxies, and their integrated
colors. New calibrations of absolute SBF magnitude in $V$, $R$, $I$, and $K$ 
photometric filters are provided, which appear reliable enough
to directly gauge distances bypassing other distance indicators.

The SBF technique is also used as stellar population tracer to derive
age and metallicity of a selected sample of galaxies of known
distances. Finally, {\it SBF color} versus {\it integrated color}
diagrams are proposed as particularly useful in removing the well
known {\it age-metallicity degeneracy} affecting
our knowledge of remote stellar systems.

\end{abstract}


\section{Introduction}

Our understanding of external galaxies, which are milestones in the
cosmological evolution, relies on the ability of gauging their
distances, ages, and chemical compositions.  Surface Brightness
Fluctuations (SBF) are recognized as a powerful tool for providing
accurate answers to all these challenges (Tonry, Ajhar \& Luppino
1990: TAL90; Buzzoni 1993: B93; Worthey 1993: W93; 
Brocato, Capaccioli \& Condelli 1998;
Blakeslee, Ajhar \& Tonry 1999; Liu, Charlot \& Graham 2000: LCG00;
Blakeslee, Vazdekis \& Ajhar 2001: BVA01). Since the first attempt of
using the SBF in galaxies to estimate cosmic distances (Tonry \&
Schneider 1988: TS88), a major observational 
effort has been made
to improve the accuracy of the method (TAL90; Luppino \& Tonry 1993;
Lorenz et al. 1993; Tonry et al. 1997: TBAD97). As a matter of fact,
the typical uncertainties on the distances derived via the SBF are of
the order of 10\% or less (Mei, Quinn \& Silva 2001). Moreover, SBF work well
in a wide range of distances, from the Local Group objects out to
galaxy clusters such as Virgo, Fornax, and even Coma. Recently Jensen
et al. (2001) succeeded in reaching galaxies at 150 Mpc ($H_0 = 75~km~
s^{-1}Mpc^{-1}$) using near-infrared (NIR) HST data.

The largest SBF database available so far is that of Tonry \&
collaborators (Tonry et al. 2001: T01, and references therein).
They measured the $I$-band SBF of nearly 300 galaxies, and were able
to calibrate empirically the absolute SBF amplitude in this same band
with reference to Cepheids-derived distances. The calibrations in
other bands have been provided: by BVA01 in $V$-band 
in the $K$-band by Liu, Graham \& Charlot (2002:
LGC02), and in the $F160W$-band by Jensen et al. (2003).
Relying upon measurements of dwarf elliptical
galaxies (e.g. Jerjen, Freeman \& Binggeli 1998; Jerjen, Binggeli \&
Freeman 2000), Jerjen et al. (2001) calibrated the $R$-band SBF
absolute magnutide versus the galaxy $(B-R)_0$ color.

This paper provides a new approach to derive SBF for Simple Stellar
Populations (SSPs, see Renzini \& Buzzoni 1986). Our first goal is to
revise the theoretical calibrations of absolute magnitudes of the
optical and NIR SBF. The reason is that empirical calibrations rest on
some other indicators, thus confining SBF to the role of at least
secondary distance indicators (SBF errors sum up to the uncertainties
of the primary distance indicators). Such a limitation may be removed
by theoretical SBF models which, by providing {\it absolute galaxy SBF
magnitude } versus {\it integrated color} relations for several different
filters, set the SBF method into the class of {\it primary distance
indicators}. Of course one needs a high degree of reliability for the
adopted theoretical framework.

A second goal is to probe the capability of our SBF models in
analyzing the stellar content of star clusters and galaxies as a
complementary tool to the {\it classical} method based on integrated
magnitudes and colors.  Unfortunately, disentangling the information
on the main physical quantities which define the dominant stellar
populations, the formation, and the evolution processes in elliptical
and spiral galaxies from their integrated broad-band colors and
spectral features still remains an intriguing problem 
(Brocato et al. 1990; Weiss, Peletier \& Matteucci 1995;
Bressan, Chiosi \& Tantalo 1996; Kobayashi
\& Arimoto 1999; Trager et al. 2000: TFWG00).  For
instance, even in the simple case of elliptical galaxies, age works in
masking possible metallicity effects (Renzini \& Buzzoni 1986; Worthey
1994). Since SBF are generated by the poissonian fluctuations of the
stellar counts in each pixel of a galaxy image, they have a well understood
physical origin (dictated by the properties of the stellar population
where they originate from). The amplitude of these fluctuations is
determined by the age, chemical composition, and distribution of the
evolving stars in the proper evolutionary phases. Thus, theoretical
SBF colors and magnitudes are expected to provide a diagnostic
of the stellar content of galaxies, useful in decoupling age from
metallicity (LCG00, BVA01).

So far, only a handful of theoretical works on the SBF modeling are
available (W93; B93; LCG00; BVA01; Mei, Quinn \& Silva 2001). In
spite of this impressive results, still models based on different
ingredients (stellar evolutionary tracks, temperature-color
transformations) and detailed assumptions (HB morphology, AGB
evolution, mass loss, etc.)  are needed in order to compare the
various theoretical expectations and thus improve the reliability of
the theoretical framework. A lightening example of the present
situation is offered by Fig.~5 of LGC02, clearly showing that the
inferred ages and metallicities depend on the choice of models.

The presentation of our work is organized as follows. The basis of the
model is described in Section 2, where we also present absolute
magnitudes and colors for the optical and NIR ground-based and HST
filters under different assumptions on the chemical composition and
the age. Section 3 reports on a comparison between our SBF predictions
and those by other authors. The effects of the Initial Mass Function
(IMF) and the contribution of stars in late evolutionary phases are
explored in Section 4, together with a discussion on the impact of
using different color-temperature libraries. After testing our SBF
models as distance indicators on Galactic Globular Clusters (GGCs) and
galaxies, in Section 5 they are applied to derive population
parameters of some galaxies. Theoretical calibrations are
presented in Section 6. The main results are summarized in Section 7.

\section{SBF synthetic model}

Our SBF model is based on the stellar population synthesis code
described and tested observationally in Brocato et al. (1999;
2000: BCPR00). This code is optimized to reproduce both the observed
distribution of stars in the color-magnitude diagrams (CMDs) and the
integrated properties (color and spectral indices, energy
distribution, etc.) of GGCs and galaxies. Here we report the main
ingredients used in the present version of the code, referring to the
quoted papers for details on the theoretical background.

All the computations refer to Simple Stellar Populations,
i.e. single-burst stellar systems with homogeneous chemical
composition, and an IMF to be randomly populated by our model machinery
(Appendix A). Those models computed by assuming a bimodal
Scalo-like IMF in the mass range $0.1-1.5$ ${\cal M}_{\sun}$ will be called
{\bf reference models} in the forthcoming.

The database of stellar evolutionary tracks is from the
Teramo-Pisa-Roma group, based on the evolutionary stellar code FRANEC
(Cassisi et al. 1998, and references therein). Here we include tracks
with metallicity ranging from $1/200~Z_{\sun}$ to super-solar
metallicity $2Z_{\sun}$ (Castellani, Chieffi \& Pulone 1991;
Castellani, Chieffi \& Straniero 1992; Bono et al. 1997a,
1997b). These sets of tracks cover all of the major evolutionary
phases: main sequence (MS), sub-giant branch (SGB), red giant branch
(RGB), He-burning phase (horizontal branch, HB), asymptotic giant
branch (AGB). The latest stages, which are crucial for SBF
calculations, have been treated with special care;
they extend to Post-AGB (PAGB) and white dwarf (WD) sequences when
needed.

In each reference model, mass loss along the RGB and
AGB is taken into account following the prescriptions by Reimers
(1975).  In particular, along the RGB we adopt $\eta=0.4$, which is
able to reproduce the typical HB stars distributions and the
integrated colors in globular clusters (Renzini \& Fusi Pecci 1988;
BCPR00). A slightly higher value is taken for stars in the upper part of
the AGB ($\eta_{AGB}=0.8$, see paragraph 4.2.2), which reproduces
the observed amount of mass loss of very bright stars (Le Bertre et
al. 2001).

The magnitudes of the stars are derived assuming the bolometric
corrections and the colors by Lejeune, Cuisinier \& Buser (1997,
LCB97) for the standard Johnson-Cousins $UBVRIJHK$ photometric system.
We provide also SBF amplitudes for a set of HST {\it WFPC2} and {\it
NICMOS} filters covering almost the same range of wavelengths (namely $F439W,
F555W, F814W, F110W, F160W$ and $F222M$), adopting the transformation
tables by Origlia \& Leitherer (2000, OL00). The effects of other
transformation recipes, resting on different assumptions, will be
discussed in Section 4.3.

As an example of the color-magnitude diagrams of the SSP
under investigation, we plot in Fig.~\ref{fig1} a sample
of CMDs for a fixed age t=15 $Gyr$, and for three chemical compositions
($Z=0.001, 0.01,$ and 0.02).

The present procedure to derive the SBF differs from the
``usual'' one, based on the ratio of the second to the first moment of
the luminosity function (TS88; W93; Worthey 1994; LCG00; BVA01). Our
method is briefly outlined here; a detailed description and a
discussion are reported in Appendix A and in Cantiello (2001).

By a Monte Carlo random method, we compute a SSP model with a fixed
number of stars $N_{star}$, for a given set of stellar population
parameters: age, chemical composition, IMF, etc.. This provides an
integrated flux (${F}_{X}$) within each specific filter $X$. The same
set of parameters is used to compute $N_{sim}$ independent
simulations. The corresponding SBF are explicitly calculated as:

\begin{equation}
\bar{F}_{X} =\frac{ \langle { F_{X} - \langle {F_{X}} \rangle }\rangle^2}{\langle F_{X} \rangle}
\end{equation}
where $\bar{F}_{X}$ is the flux fluctuation from which we derive the
SBF amplitude $\bar{M}_{X}$.

To match the typical number of stars per pixel in real observations
$N_{star}$ is chosen to be $\simeq 10^{5}$ (e.g. B93).  By increasing
$N_{sim}$ (to a maximum of $10^5$ experiments), the ``fluctuations'' of
$\bar{M}_{X}$ tend to vanish, i.e. $\bar{M}_{X}$ tends to an
asymptotic value.  As an example, Fig.~\ref{fig2} shows the value of
the $I$-band SBF amplitude ($\bar{M_{I}}$) versus $N_{sim}$ for a
model with $Z=0.02$, $Y=0.289$ and $t=15~ Gyr$.  For small $N_{sim}$
(say, $N_{sim} \lsim$500), $\bar{M_{I}}$ oscillates strongly.  For
$N_{sim} > 1000$ simulations, $\bar{M_{I}}$ approaches the asymptotic
value $\bar{M}_{I}^{asym}\simeq -1.47~mag$ and keeps well within the
best observational accuracy of SBF measurements ($\sim$0.05,
T01), i.e.  $|{\bar{M_{I}}-\bar{M}_{I}^{asym}}|\lsim$0.05.
By repeating the same procedure for different sets of the parameters
of stellar populations and for different filters, we found that
$N_{sim}^{*}=5000$ is appropriate to all the photometric bands used
here for the adopted $N_{star}\simeq 10^5$.

\subsection{SBF versus Chemical composition and age }

In order to cover the typical chemical compositions and ages of both
GGCs and galaxies, we computed the SBF absolute magnitudes for
the following set of chemical compositions: ($Z=0.0001, Y=0.23$);
($Z=0.001, Y=0.23$); ($Z=0.01, Y=0.255$); ($Z=0.02, Y=0.289$); and
($Z=0.04, Y=0.34$).  The values for the ages are $t=5, 9, 11, 13$, and
$15~ Gyr$.  The resulting SBF are listed in Table~1 $(a,b)$ for
ground-based and HST filters. The complete output of the models
(CMDs, integrated colors, and SBF amplitudes) are available on
request via web.

Fig.~\ref{fig3} reports the SBF predictions as a function of
metallicity. We consider two ages, $t=15$ and $t=5~Gyr$, and both
sets of filters (Fig.~\ref{fig3}a refers to ground-based, Fig.~\ref{fig3}b
to HST filters). In the optical range all the SBF amplitudes become
fainter with increasing metallicity, while the opposite trend is shown
by NIR bands. This behavior does not depend on the age, at least in
the range investigated in this paper. Moreover, it supports the
conclusion by W93 of the possible existence of an ideal filter
centered at $\sim1\mu m$ that is expected to be independent of
metallicity.

For $Z\lsim0.01$ the least sensitive band is $I$; this means that
$\bar{M}_{I}$ is an excellent distance indicator for metal poor old
stellar systems.  At higher metallicities ($Z\gsim0.01$) the slope of
the relation $\bar{M}_{X}$ versus $Z$ increases for all the SBF
amplitudes and is similar for all bands ($BVRI$) but for $U$.  In
general, $\bar{M}_{U}$ has a quite different behavior in the whole
range of metallicity, varying of $\sim 3.2~ mag$ due to the strong
decrease of the $U$ mean luminosity of the brightest stars in the
population as the metallicity increases (Fig.~\ref{fig1}). Although
ruling out $\bar{M}_{U}$ as a reliable distance indicator, this high
sensitivity makes it an excellent population tracer. Unfortunately, no
SBF measurements are available in the $U$ filter so far.  The
predicted small variations of $JHK$ bands at $Z\gsim0.01$, coupled
with high luminosity, support the increasing number of SBF
measurements in these bands.  The SBF amplitudes computed for the HST
filters (Fig.~\ref{fig3}b) show a behavior quite similar to the
corresponding ground-based filters.

The next step is to understand the effects of variations in the age of
the stellar population on the SBF amplitudes.  The results of our
computations for models of age $t=$5, 9, 11, 13, 15 Gyr are plotted in
Fig.~\ref{fig4} for both a metal-rich ($Z=0.02$), and a metal-poor
($Z=0.0001$) stellar system. In general, the sensitivity to age
grows from $U$ to $I$, with changes remaining below $\sim0.4~
mag$. $JHK$ bands are more sensitive; SBF amplitudes increase by about
$0.7~mag$ from 5 to $15~ Gyr$, in agreement with BVA01 and
LCG00. However, for stellar populations with age in the range under
consideration, $JHK$ SBF are strictly dependent on the adopted
treatment of the brightest evolutionary phases (thermal pulses, mass
loss processes, etc.), as we will show in Section 4.2.

Let us finally remark that the SBF amplitudes for old stellar
populations are much more sensitive to metallicity changes than to
age, particularly at optical wavelengths. A similar conclusion, even
with slightly different absolute SBF values, has been reached by other
authors (B93; LCG00; BVA01), who adopt different approaches and
ingredients to derive population synthesis models.  In this sense this
result may be regarded as theoretically well established, i.e. it does
not depend on the adopted stellar population models.

\section {Comparisons with other models}

We compare our work with the models by BVA01 and LGC02.  Since
the grids of ages and metallicities do not match, for the comparison
we select the values as similar as possible. We restrict the
discussion to populations of fixed age ($t\sim 15~Gyr$), and to $V,I$,
$F160W$ photometric bands. Fig.~\ref{fig5} plots the SBF
predictions versus the $(V-I)_0$ ({\it Right panel}) and $[Fe/H]$
({\it Left panel}) for single-burst stellar population models.

In spite of the very different theoretical framework, the general
trends of the SBF amplitudes versus both metallicity and the integrated
$(V-I)_0$ color are the same for all authors.  However, for metallicity
higher than solar, BVA01 and LGC02 predict fainter SBF amplitudes in
$V$ and $I$ bands (up to $\sim0.5~mag$) and a similar, but less
relevant, difference appears in the $F160W$-band.

The reasons for such a discrepancy are expected to be in the
basic ingredients and in the transformation tables. Our models are
based on a completely different evolutionary framework compared to
BVA01 and LCG00, whose models rely on the Padua stellar evolution
database, although they use two different compilations of isochrones
(Bertelli et al. 1994, and Girardi et al. 2000 respectively), and
different prescriptions for the AGB stars.  Furthermore, we adopt a
different approach to populate the brightest evolutionary phases
(HB, AGB, thermal pulses) as we shall show in the following
Sections. For an extensive discussion on the impact of evolutionary
tracks and transformations tables on SBF expectations, the reader may
refer to LCG00 and to our Section 4.3.

Let us remark that there is a substantial difference in the
meaning of the two panels of Fig.~\ref{fig5}: $[Fe/H]$ is an
input parameter for all the models, so it is directly comparable,
while the $(V-I)_0$ color is an output of each code, then it depends
on the adopted evolutionary tracks, on the procedure in populating the
isochrones, and so on.  Obviously, it may vary from model to model
even for the same input parameters.

\section{On the assumptions of the reference models}

Starting from the reference models, in this Section we investigate the
sensitivity of SBF to IMF, late evolutionary phases, and
color-temperatures relations.

\subsection{Initial Mass Function}

Usually the mass spectrum of a stellar population is described by the
simple power law proposed by Salpeter (1955) or by the bimodal (or
more) function of Scalo (1986 and thereafter). Indeed, a crude
extrapolation of the Salpeter law to masses ${\cal M} \lsim 1 {\cal
M}_{\sun}$ seems to fail, while the flattening and even possible
declining behavior of the Scalo IMF appears more appropriate. For
these reasons, in our reference calculations we have adopted a
Scalo-like IMF. Nevertheless the slope of this function at the lowest
masses still remains very uncertain (Reid 1998). In this Section we
evaluate the contribution to the SBF by the lowest mass stars, and
what happens under significant changes in the slope of the IMF. Two
different numerical experiments are performed on an old
solar-metallicity SSP ($t=15~Gyr$, $Z=0.02$, $Y=0.289$):

$a)$ SBF amplitudes are computed for a set of models obtained by
adopting an IMF starting from different low mass limits, namely
${\cal M}_{low}$=0.85, 0.6, 0.4, and 0.1 ${\cal M}_{\sun}$, by keeping 
constant the
number of RGB stars (actually $\sim 100$). For all these simulations
we adopt a Salpeter-like slope ($x=1.35$) in order to enhance the
contribution of very low mass stars (Table~2).

The table shows that the SBF amplitudes increase with the mass limit
${\cal M}_{low}$, the differences being fairly negligible or small in
all cases but in the interval $0.6-0.85 {\cal M}_{\sun}$, whose upper
limit is of the order of our Turn-Off Mass (${\cal M}_{TO}$, in the
model quoted here ${\cal M}_{TO}\simeq 0.90 {\cal M}_{\sun}$). As a
matter of fact, the ignorance of the upper part of the MS 
(${\cal M}_{TO} - {\cal M}\leq 0.25 {\cal M}_{\sun}$) leads to underestimate
the total luminosity of the population, which is obviously relevant in
computing SBF because it represents the denominator of eq.~1.  On the
contrary, to add MS stars with ${\cal M} < 0.6~{\cal M}_{\sun}$ causes
small variations of the SBF, since the total luminosity of the stellar
population does not change significantly. Note that this is true even
though stars with ${\cal M} < 0.6 {\cal M}_{\sun}$ represent nearly
90\% of the total number of stars in the population. In conclusion, no
matter how (reasonable ways) the mass spectrum at 
${\cal M} < 0.6~{\cal M}_{\sun}$ is chosen, its contribution to SBF will not
exceed $0.1~ mag$ in the most sensitive band $K$.

$b)$ The effect on the SBF values due to the IMF slope is explored by
varying the exponent of the power law ($x$=0.35, 1.35, and 2.35) for
masses ${\cal M} > 0.4~ {\cal M}_{\sun}$ and assuming a flat
distribution for masses ${\cal M} < 0.4~ {\cal M}_{\sun}$
(Table~3). Not surprisingly the variations are rather small 
($\Delta \bar{M}_{X} \lsim 0.2~ mag$); $x$ affects mainly the number of the MS
stars, which in turn affect only slightly the predicted SBF
amplitudes, as previously shown. However, an increased/decreased
number of post-MS stars is foreseen if the exponent has a lower/larger
value than the Salpeter's one. Since the fluctuation rests
mainly on the RGB stars, this explains why SBF brighten with
decreasing IMF exponents, and viceversa.

We conclude that variations on the IMF slope and/or the contribution
of low mass stars to the SBF amplitude, {\it although not negligible},
play a secondary role with respect to the quantities (and assumptions)
affecting the post-MS evolution, as for example the chemical
composition and age of the populations.
The result obtained with our new and fully independent
code strengthens what originally found by W93, and recently
confirmed by other authors as well (e.g., BVA01).

\subsection{SBF versus brightest stars}

As just recalled, the SBF signal is very sensitive to the properties
of the most luminous stars of the stellar population, as it is the
ratio of the second to the first moment of the stellar luminosity
function. This requires the SSP models to reach a high level of
accuracy in reproducing the post-MS evolution. Therefore, one of the
aims in optimizing our code has been to fine-tune the stellar counts
during the latest and fastest phases experienced by low mass stars
along their evolution (see also BCPR00).

Below we discuss the expected SBF sensitivity to the very bright
stars in post-MS evolutionary phases: HB morphology, thermal pulses
(TP), and PAGB phases.

\subsubsection {Horizontal Branch Morphology}

As well known, the main parameter regulating the HB morphology of a
generic stellar population is the metallicity (Lee, Demarque \& Zinn
1987).  If one considers just the effect of $Z$, the expected HB
morphology is that of Fig.~\ref{fig1}: {\it blue} or {\it
intermediate} for $Z\lsim$0.001, {\it red} when $Z > 0.004$. Thus, the
distribution in color of low mass stars during the central He-burning
stage is expected to play a major role in predicting $UB$ SBF (see
also W93).  This is particularly true for high metallicities since the
HB stars become brighter than the RGB tip (Fig.~\ref{fig1}).

Although we shall not discuss the nature of the second parameter
(Freeman \& Norris 1981), our models take into account that the
metallicity is not the only quantity which can affect the HB
morphology. Here, we adopt the HB morphology as directly produced by
metallicity and by mass loss along the RGB (i.e. we use the mass loss
to simulate the effect of the second parameter whatever its real
nature is). Along the same vein, one may recall that mass loss during the
RGB evolution controls the actual total mass of the HB star and then
affects its location in the CMD (Rood 1973). Moving from low to high values of
mass loss, the total mass of the HB stars decreases; this fact
implies, on average, higher temperatures and bluer colors.  On the
observational side, there are indications of mass loss enhancement for
increasing values of metallicity (e.g. Kudritzki, Pauldrach \& Puls
1987).

On this basis, we have investigated the importance of HB
morphology on SBF by simulating different amounts of mass-loss along the RGB,
i.e. we have changed $\eta$ in the Reimers formula ($\eta=0.2, 0.4,
0.6$, and $0.8$) in computing a set of SSP models with given
metallicity ($Z=0.01, 0.02$, and $0.04$) and age ($t=15~Gyr$).

Table 4 summarizes the SBF values for all ground-based filters. The
effects on $UVIK$ bands SBF amplitudes are shown in
Fig.~\ref{fig6}. The growth in luminosity of $U$-band SBF and the
constancy of the other optical bands confirms the above expectation.
We note that the effect is more pronounced for $Z=0.01$ ($\Delta
\bar{M}_{U} \sim2~ mag$).  For this metallicity, an enhancement of
the mass loss on the RGB reduces the mass of the ``forthcoming'' HB
stars by a quantity large enough to populate the blue side of the HB.
The larger is the mass loss, the larger is the number of blue HB stars
and, consequently, the larger is the $U$-band SBF amplitude.  For
higher metallicity systems, variations are still present but of
smaller amplitude ($\Delta \bar{M}_{U} \sim 0.5~mag$) just because
the mass of the RGB stars for the selected age is so large that only
red HB stars are produced even for the largest mass loss rates
considered here.

From Fig.~\ref{fig6} it is also apparent the opposite behavior shown by
$\bar{M}_{K}$ which depends on the fact that, as $\eta$ increases, the
number of bright stars evolving on the AGB is smaller.

Let us note here that an age of $15$ $Gyr$ is selected because
models with younger ages provide a red clumpy HB morphology
independently of any reasonable value of the $\eta$ parameter. This
is because, by decreasing the age, the HB is populated by stars of
increasing mass, which spend their He-burning phase at the red portion
of the HB.  Thus, no changes are expected on SBF.  An unlikely
value of $\eta \sim 2.0$ along the RGB phase would be needed to force
the models to a blue HB morphology. However, this extremely high mass
loss on the RGB would imply a similar (or larger) efficiency along the
AGB. This means that the stars are expected to lose their H-envelope,
and quickly move to the WD cooling sequence during the very early part
of the AGB ascension.  Such an improbable scenario leads to SBF which
are found to be fainter, particularly for the redder color (up to
$1~mag$ for the $K$-band), when compared to models with more realistic
assumption on the RGB mass loss efficiency. 

The present results, even if they do not provide an exhaustive picture
of the problem, give quantitative indications on the $U$ and $B$ bands SBF
variations as a function of the HB morphology. The effect of $\eta$
on SBF was also explored by B93, but his synthetic models have always
a red clumpy HB morphology for populations with $Z>0.001$.

We conclude that changes in the HB morphology, as induced by
reasonable assumptions on the mass loss along the RGB, have a
negligible impact on SBF, if one considers filters redder than $B$.

\subsubsection {Thermal Pulses}

Mass loss strongly affects the He-exhaustion stages too, and determines
different scenarios for the evolution of stars leaving the HB, either
to the TP phase or rapidly to a PAGB
condition. According to the observations (e.g. Alard et al. 2001),
mass loss becomes more and more efficient as the star climbs the AGB,
till the ejection of the stellar envelope during the last superwind
stages (Renzini 1981).

To extend our He-burning evolutionary tracks to the TP phase, we
follow the prescriptions by Groenewegen \& de Jong (1993) as in a
previous paper (BCPR00).  As cited above, our reference models
simulate the mass loss process along the AGB by adopting an enhanced
Reimers coefficient $\eta$ (namely $\eta_{AGB}=0.8$).  The
corresponding mass lost by stars is in the range from $10^{-7}$ to
$10^{-6} {\cal M} _{\sun} ~ yr^{-1}$, in fair agreement with observations of
AGB stars (e.g. Le Bertre et al. 2001).

If the age of the population is fixed, an increase of the mass loss
rate reflects into a decreasing of the number of very bright AGB stars
just because their H-rich envelopes become quickly so thin that the
stars leave the AGB phase. According to this simple view, an increase
of the mass loss rate induces a corresponding decrease of the foreseen
luminosity of the AGB tip. This occurrence has to affect SBF
amplitudes. To investigate such an issue quantitatively, we computed a
series of SBF models by changing $\eta_{AGB}$ from 0.4 to 3.2. The
results in the Johnson-Cousins filters 
are reported in Table~5 for solar metallicity and
the labeled ages. Dealing with AGB stars, the effect is more
remarkable in the redder bands.  Fig.~\ref{fig7} illustrates the SBF 
dependence on $\eta_{AGB}$ for models with different ages and solar
chemical composition in the $IJHK$ filters.

Such a dependence provides the unique opportunity of using the SBF to
constraint evolutionary parameters during the late stage of the AGB
and TP phase. In particular, we calibrate the mass loss efficiency
along the AGB.  Each panel of Fig.~\ref{fig8} shows the $I$-band SBF
predictions computed with a specific value of $\eta_{AGB}$
(0.4, 0.8, 1.6, 3.2). To
perform this calibration we use the mean empirical relations
$\bar{M}_{I}$ versus $(V-I)_0$ as derived from observations of galaxies 
and GGCs.  
The inclined solid line in the range $1.0 \lsim (V-I)_0 \lsim 1.3$
represents the relation derived by T01.
According to T01 the slope of this relation is quite well
established, whereas the zero point is more uncertain with a rms
scatter as large as $0.1~ mag$ (see Table 4 in TBAD97).  However, the
true indetermination of the zero point is expected to be of the order
of $0.20~mag$, since one has to include the uncertainties in the
Cepheid zero point (Mould et al. 2000, Freedman et al. 2001) and in
the LMC distance modulus (e.g. Benedict et al. 2002).  The horizontal
solid line in the range $0.8 \lsim (V-I)_0 \lsim 1.0$ refers 
to the observed locus
of GGCs (Ajhar \& Tonry 1994: AT94), after a recalibration according
to BVA01 prescriptions (eq.~5 in BVA01).  In this case the quoted rms
scatter is $\sim0.25~mag$.  In addition, the zero point of the
horizontal line has an indetermination of at least $0.1-0.15~mag$
arising from the uncertainty of the GGCs distances.

Fig.~\ref{fig8} shows that the two extreme values of the $\eta_{AGB}$
do not match the observed SBF. In fact $\eta_{AGB} = 0.4$ provides too
bright SBF amplitudes at all metallicities but $Z = 2Z_{\sun}$; on the
contrary, the models with $\eta_{AGB} = 3.2$ appear definitively
fainter than observations. The intermediate cases are more
interesting: 
both the $\eta_{AGB}$ values fairly reproduce the SBF data. 
However, the horizontal line
representing the GGCs measurements is fitted by SBF simulations of the
proper metallicity values but of quite different ages. The assumption
of $\eta_{AGB} = 0.8$ leads to a remarkable fit with models of
$13-15~ Gyr$, while $\eta_{AGB} = 1.6$ requires an age of the order
of $\sim 9~ Gyr$. Since our SSP models reproduce the CMDs and
integrated properties of GGCs when an age of $\sim 15~ Gyr$ is
assumed (BCPR00), we find that $\eta_{AGB} = 0.8$ simulates the mass
loss process on AGB in such a way that the resulting distribution of
stars on the AGB is able to reproduce at the same time the SBF
measurements in galaxies and in GGCs.

In spite of the large uncertainties on the zero points, this 
result represents a relevant constraint
for stellar evolution models of AGB stars.
Even if the exact values of $\eta_{AGB}$ depend
on the assumed theoretical framework, the method outlined here remains
a new and powerful tool to improve our knowledge on the details of
the evolution of AGB stars.

\subsubsection {Post-AGB stars}

The theory predicts a small number of PAGB stars (Renzini 1998,
BCPR00) which, however, increases with the total number of stars. 
Though rare objects, PAGB stars cannot be ignored when dealing
with populous systems such as galaxies (Greggio \& Renzini 1990; Brocato et
al. 1990).

For the complex evolution off the asymptotic branch we rely on the
stellar models by Vassiliadis \& Wood (1994, VW94) and Bl\"ocker \&
Sch\"omberner (1997, BS97).  Since SBF are sensitive to the most
luminous stars, as first approximation, we simulate only the most
luminous PAGB, neglecting Post-Early-AGB evolved stars, which leave
the Hayashi track at lower luminosity before the first thermal pulse
(Castellani \& Tornamb\`e 1991). Under these assumptions, we computed
SBF amplitudes for the reference stellar populations with solar
metallicity and $t=15~Gyr$, by simulating a different amount of PAGB
stars ($N_{PAGB}$) for the same total number of stars.  To perform
these simulations, firstly we evaluate the number of PAGB with respect
to the number of expected HB stars ($N_{HB}$) by following the
prescriptions by BCPR00. Secondly, we explored a wide range of values
of $N_{PAGB}$/$N_{HB}$ by artificially varying this ratio.

In Fig.~\ref{fig9} we plot the $UBV$ and $I$ bands 
SBF amplitudes versus the
ratio $N_{PAGB}$/$N_{HB}$. The shadowed region corresponds to the
value of $N_{PAGB}$/$N_{HB}$ properly derived according to BCPR00.  It
is apparent that SBF obtained in filters redder than $V$ are not
affected by the presence of bright PAGB stars. On the other hand, $U$
and $B$ SBF appear to be quite sensitive to the number of predicted
PAGB stars, independently of the assumed stellar evolutionary tracks
(VW94, BS97). Here, we suggest that precise measurements of $U$ and
$B$ SBF, with telescopes having a high efficiency at such wavelength
(for example the VLT Survey Telescope, or VST: 
http://twg.na.astro.it/vst/vst\_homepage\_twg.html), would provide
the unique chances of deriving valuable constraints to stellar
evolution theory of PAGB stars.

Finally, we find that the luminous and hot stars crossing the
CMD from the AGB tip to the beginning of the WD cooling sequence
do not appreciably affect the SBF amplitude in most of the bands but
$U$ and $B$. We also note that the stars evolving down along
their WD cooling sequence are not expected to provide a significant
contribution in optical bands. In fact, it 
becomes more and more
negligible as the WDs luminosity decreases, even though they pile up due
to longer evolutionary times.

\subsection{Impact of Atmospheres transformations}

The relations transforming the theoretical quantities
($log~L/L_{\sun}$ and $log~T_{eff}$) into observables (magnitude
and colors) are of fundamental importance in modeling SBF. In
general, theoretical relations between colors/bolometric constant and
temperatures suffer from the complexity in treating molecular and
atomic transitions when the metal content increases and the star
temperature decreases. When the metal content grows up to solar or
super-solar values, these difficulties become more and more severe due
to the formation of grains and complex molecules in the outer part of
the atmosphere, which cause strong blanketing effects (Bell 1973). On
the other hand, empirical compilations involve difficulties in
deriving fundamental star parameters as metallicity and gravity from
observed spectra and colors (Alonso, Arribas \& Martinez-Roger, 1996,
1999).

In this Section we evaluate how SBF amplitudes change by adopting
different transformation tables available in literature. To provide a
quantitative comparison we selected three color-temperature
compilations commonly used for synthetic stellar population models:
the theoretical tabulation by Castelli, Gratton \& Kurucz (1997:
CGK97), the hybrid library by LCB97, and the semi-empirical Yale
transformations (Green et al. 1987: G87).  The CGK97 tabulation is the
result of the atmosphere models obtained with the updated Kurucz's
code.  LCB97 corrected tabulations are based on the theoretical
spectra from different authors, properly adapted to match observations
(see LCB97 for more details). The semi-empirical tables proposed by
the Yale group result from the calibration of the old Kurucz models
(1979). Both CGK97 and LCB97 include all the ground-based
Johnson-Cousins filters ($UBVRIJHK$), while G87 tables cover only the
optical range ($UBVRI$).

We computed a set of stellar population models for the selected
metallicities, $t=15~ Gyr$, and with the reference
assumptions on mass loss and IMF.  For each model we evaluated SBF
amplitudes by adopting all the three tabulations mentioned above.

The results are reported in Table~6. The relative differences in
the SBF amplitudes for selected filters ($UBVRIJHK$) are plotted in
Fig.~\ref{fig10} as a function of the population metal content, $Z$,
for a fixed age. The differences are computed relatively to LCB97.
For $Z<0.01$ Fig.~\ref{fig10} shows that SBF amplitudes are quite well
determined. In this metallicity range (GGCs) there are differences
$\lsim0.2~mag$ by using either CGK97, or LCB97, or G87; a fact
supporting the reliability of SBF models of these metallicities.

The situation worsens for higher metallicities. For $Z>0.01$, G87
transformations lead to systematically brighter SBF amplitudes in all
the considered bands (up to more than $1~mag$ for the $I$-band).
Such a large difference may be due to the fact that G87
transformations have been calibrated with GGCs, which are low
metallicity objects.  LCB97 and CGK97 models give quite similar
predictions for all the optical and NIR bands.  Sizeable differences
are only expected in $V$ and $R$ bands, where the SBF are about $0.3~
mag$ brighter in CGK97 models.

In order to understand the origin of these differences we show the
color-magnitude location of the most luminous stars of a
solar-metallicity isochrone aged $15~ Gyr$ transformed by adopting all
the quoted compilations. For clarity, we focus the attention on RGB
stars, bearing in mind that AGB stars behave similarly. We consider as
representative the ($M_V$, $(V-I)_0$), ($M_I$, $(V-I)_0$), and ($M_K$,
$(V-I)_0$) diagrams in Fig.~\ref{fig11}.  The isochrone mapped in the
observational plane by using G87 provides a much brighter RGB tip
(where stars reach temperatures lower than 3500 K) in the $I$-band
(Fig.~\ref{fig11} middle panel) with respect to LCB97 (and CGK97);
this could explain the brighter $I$-band SBF amplitude obtained by
using G87. Moreover, since the tip magnitude calculated with LCB97 is
fainter than those with CGK97 for all bands, one expects SBF to become
fainter and colors redder for metallicity higher than solar when LCB97
is adopted, as exactly shown by Fig\ref{fig10}.

In Fig.~\ref{fig11} we also plot the isochrone transformed with the
empirical temperature-color relations published by Alonso et
al. (1996, 1999), since they are used in BVA01 SBF models. 
We note that $RI$ filters involved in the Alonso et al. transformations
are the Johnson \& Morgan (1953) bands, with respect to the Cousins (1976,
1978) passbands commonly used for SBF observations. By applying the relations
to convert $(V-I_{J})$ into $(V-I_{C})$ (in
Fig.~\ref{fig11} we used Bessel 1979), the differences become
negligible, making Alonso et al. models similar to the others in their
own temperature range. A significant disadvantage of this grid is
that it is limited to $T_{eff} \geq 3750$ K for
$[Fe/H]>-1.0$, then one needs to extrapolate or to merge with other
tables down to lower temperatures. For all these reasons we decided
not to use these color-temperature relations in computing SBF.

\section{Comparison with the observational data}

As a first step we compare our SBF predictions in several bands with
actual observations of stellar systems (GGCs and galaxies) with known
distances. Then, the theoretical SBF are used to derive the
characteristics of a selected sample of observed stellar populations.

\subsection{Galactic Globular Clusters }

The only SBF measurements related to GGCs are those by AT94 for a
sample of 19 objects. There are several reasons to study the SBF of
these objects: 1) GGCs span a range in metallicity down to values
lower than expected in galaxies; 2) since they are thought to be
formed by the gravitational collapse of a homogeneous gas cloud, they
are the best observational counterpart of theoretical SSPs; 3) the SBF
of GGCs are determined directly from their resolved stars (differently
from galaxies, which require a more complex analysis). Thus GGCs SBF
represent a simple and unambiguous target to verify the stellar
evolution assumptions.

Before going on, it is worth noticing that, since the amount of
stars in a globular cluster ($10^{5}-10^{6}$) is far lower than
the stellar content in a typical galaxy ($> 10^{10}$), this might
bear on the SBF evaluations due to statistical reasons.
Consequently, the proper way to evaluate theoretical SBF for GGCs
is to compute simulated CMDs (with total number of stars
$N_{star}\sim 10^{6}$) and apply directly the equations $7-9$ of
TS88. In spite of the difference in the total number of
stars, the resulting SBF are quite similar (within the uncertainties) to
the values for $Z\leq0.006$ 
derived from the procedure presented in Section 1
(Table~1). Such a stability in the computed SBF amplitudes is due
to the compensation of two conflicting effects, which arise when
the size of the simulated stellar population decreases. On one
side, the number of stars expected in the fastest and brightest
evolutionary phases (e.g. TP, AGB) tends to vanish, and fainter
SBF amplitudes are obtained. On the other side, the number of
bright RGB stars decreases providing less populated RGB;
statistical effect become relevant (see also BCPR00), leading to
brighter SBF amplitudes. 

Since AT94 measurements are available only for $VI$ bands, the
comparison with theoretical models can be done only for these two
filters. In Fig.~\ref{fig12} the GGCs observational data are
compared with our models.  The observational absolute SBF magnitudes
for each globular cluster are obtained from the AT94 data according to
the expression: $\bar{M}_{X}=\bar{m}_{X}-DM$, where DM is the
distance modulus.  The latter quantity, the metallicity, and other
properties are from the latest version of the electronic catalog
compiled by Harris (1999), and are summarized in Table~7, which also
reports the observed SBF values.  In the figure the solid
(long-dashed) line connect the oldest (youngest) models with different
metallicity.  The dotted lines represent the mean line according to
BVA01 prescriptions (their eq.~5).  From Fig.~\ref{fig12} it is
apparent that:

$i)$ 
the DM directly derived from the theoretical SBF are in very good
agreement with the recent evaluation from other distance indicators;

$ii)$
 the general trend of observed SBF versus metallicity is well
reproduced by theoretical models; this fact is noteworthy because the
$[Fe/H]$ values of GGCs are measured directly and not just derived by
integrated colors.

Finally, let us note that the good agreement between the single
measurements and the models is particularly relevant considering that
no effort has been put in reproducing the peculiarities of each
individual GGC (e.g. HB morphology, brightest stars, etc.).

\subsection {The Galaxy sample}

Unfortunately, no GGC with $Z \gsim Z_{\sun}$ is known, and one
needs to look at elliptical galaxies and bulges of spirals to verify
the degree of accuracy of SBF predictions for metal rich populations.

For this purpose, we extract a subsample of galaxies from the T01
$I$-band database by selecting objects with reliable distance
estimates as derived by Ferrarese et al. (2000).
The result is given in Table~8, where data of 31 galaxies
are summarized. In the table we
report from left to right: the galaxy name, the morphological type T,
the visual absorption
$A_{V}$, the integrated color $(V-I)_0$, and the SBF apparent
magnitudes in $I, V, R, K', K_{s}, F814W$, and $F160W$ filters. The
last column reports the galaxy distance derived from Cepheids,
when available (8 galaxies), otherwise the average distance from
other indicators (Ferrarese et al. 2000).
We note that, for the sample of galaxies with Cepheid-based distances, 
the new calibrations of the Period-Luminosity (PL) 
and Period-Luminosity-Metallicity (PLZ) relations by 
Freedman et al. (2001) lead 
to DM which differ on average of $\Delta DM_{PL}=0.13 \pm
0.06$ and $\Delta DM_{PLZ}=0.03 \pm 0.05$, respectively. 
Being the debate on PL
and PLZ still open, we adopt the Ferrarese et al.
database for sake of homogeneity and clearness.

We test the consistency of our predicted SBF starting from galaxies
with distances evaluations based on Cepheids, as they are the most
reliable distance indicator. Since the SBF method works at its
best with spheroidal galaxies, there is a relative scarcity of SBF
measurements for spirals, and even less objects with both SBF and
Cepheids distance determinations. We first focus the attention to the
$I$-band, due to its importance for the calibration of the empirical
relation (T01), and then we extend the analysis to other photometric
bands. Fig.~\ref{fig13} illustrates the comparison of the sampled
galaxies with the present models; the empirical line by T01 is also
plotted. The figure shows that all the galaxies well match the
theoretical models with $(V-I)_0 > 1.0~ mag$. We emphasize that the
zero point of the empirical relation by T01 is derived directly from
Cepheids-calibrated distances, while the present theoretical SBF are
completely independent of them.  At $(V-I)_0<1.0$ the theory predicts
a nearly constant value of $I$-band SBF for models of similar age, as
already verified on GGCs measurements. Moreover, this appears in
agreement with the location of the galaxy NGC 5253 (not plotted in the
figure) reported in Fig.~8 by BVA01.

By extending the comparison to all the galaxies in Table~8
(Fig.~\ref{fig14}), the agreement remains good, even if few of the
observed values appear fainter than those predicted by the theory and
by the empirical line.  We note that the galaxies showing the largest
scatters (NGC 4278 and NGC 4565) have distance measurements with a
quite large spread, if compared to internal uncertainties (Ferrarese
et al. 2000).

Taking advantage from multi-band observations, in
Fig.~\ref{fig15}a,b,c we report a similar comparison for $V,K$, and
$F160W$ filters.  Following Jensen et al. (2001) we discard NGC 4536
from their sample because its anomalous $F160W$-band SBF magnitude
appears to be due to clumpy dust contamination. Fig.~\ref{fig15} shows
that our theoretical predictions are consistent with observations not
only in the $I$-band, but also in other selected optical and NIR
filters.  This result is relevant for two reasons. First, the
agreement of absolute values of the SBF amplitudes suggests that the
present theoretical SBF, computed in the other selected photometric
bands, may be adopted to derive distances too. Second, the 
agreement over a wide range in wavelength supports the
reliability of the global theoretical assumptions on the stellar
population from which we derive the SBF.

Being the proposed theoretical framework supported by SBF
observations, at least in the main relevant features, we are led to
push forward the capability of the SBF technique and to tentatively
infer information on age and metallicity from the SBF measurements of
the quoted sample of galaxies.

\subsection {SBF as Stellar Population tracers}

Once the distance of an object is known or in the case of several
objects all at the same distance (e.g. cluster galaxies), one may want
to use the SBF models for deriving ages and metallicities of the
stellar populations.  To this end we test here the internal consistency
of the SBF models in various bands, basically $V,I,K$
and $F160W$. As further point we verify the reliability of the
inferred (average) age and metallicity through the comparison with
literature values derived by other methods.

The problem of evaluating the age and metallicity of stellar
populations in elliptical galaxies is a largely debated topic (see for
example Henry \& Worthey 1999 and reference therein). Thus, to select
measurements or evaluations of those quantities which can probe the
reliability of SBF models in deriving $t$ and $Z$ is a quite a
difficult step. For example, one has to face with the problem that
ellipticals are constituted by complex stellar populations, i.e. by
age/metallicity mixed populations (e.g.  Baugh, Cole, Frenk 1996;
Kauffmann \& Charlot 1998; Saglia et al.  2002), while we use the
approximation of single-burst stellar population. Another difficulty
comes from the evidence that metallicity and/or age gradients are
observed in some ellipticals (Peletier et al. 1990; Carollo, Danziger
\& Buson 1993, Kobayashi \& Arimoto 1999). As a further complication
SBF measurements and $t,Z$ evaluations from other indicators may refer
to not-overlapping regions of the studied galaxy.

Keeping in mind the quoted (relevant) warnings, it is still
interesting to evaluate ages and metallicities from SBF measurements
by using the present theoretical models. Clearly, the following
results refer to SBF measured in relatively extended galaxy regions,
and they are valid in the approximation that in these regions there is
a dominant stellar population (Ferreras, Charlot \& Silk 1999).  For
all these reasons the comparison between our predictions on $t,Z$, and
those obtained using other methods have to be considered with caution,
and within the quoted approximations.  The age and metallicity we
derive have large uncertainties as expected from the quoted
limitations. A more detailed investigation is out of the aim of this
paper and it should rely on mixed populations models, and also on SBF
measurements for which the radial distribution is known (e.g. Sodemann
\& Thomsen 1995).

In order to test our ability of predicting age and metallicity, we
built a sample of 15 galaxies for which these quantities have been
given by other authors (Table~9). Most of these objects come from the
compilation by TFWG00 and by Terlevich \& Forbes (2002: TF02), who
used $H_\beta$, {\it Mg b}, and $\langle Fe \rangle$ line strengths to
derive age and metallicity for a large sample of elliptical galaxies
(see the quoted papers for more details).  These estimations are
derived from observational data taken at smaller radii with respect to
SBF measurements (except in the case of Jensen et al. 2001, 2003).  In
addition, TFWG00 find indications of age and metallicity gradients
(see their Tables 6A and 6B).

To start with, let us to use the model sequences in the $\bar{M}_{X}$
versus $(V-I)_0$ plane as already shown in Fig.~\ref{fig14} and
Fig.~\ref{fig15}. The clear separation between models with low
($Z=0.0001, 0.001$) and high metallicity ($Z=0.01, 0.02, 0.04$) for
all the explored values of age confirms that this kind of diagrams are
a powerful tool for this purpose (see also BVA01, LCG00).
Since we are dealing with absolute SBF amplitudes, a
particular care has to be taken in selecting the distance moduli
required to translate the observed apparent SBF values. Obviously,
Cepheids distances are preferred; when they are not available
for a single galaxy, we use group-averaged distances. To avoid
circular arguments, we recomputed the group distance moduli by
Ferrarese et al. (2000) excluding SBF-derived distances, i.e.
averaging on the Cepheids, PNLF and TRGB distances.
The results are reported in Table~9 for the 15 galaxies.
As already stated, Freedman et al. (2001) revised the Cepheids distance scale, 
and Ciardullo et al. (2002) recently recalibrated the zero point of the PNLF.
These new results decrease the group-averaged distances of about 
$0.14~mag$ if the
Cepheid PL relation is used, but they leave nearly unchanged the above
determinations when the PLZ is considered.

Fig.~\ref{fig16} displays the theoretical models and the observational
data in selected photometric bands. As a first general result, we
find that the location of the bulk of galaxies is almost consistent
in all the photometric bands. Then we focus on two well
studied galaxies: NGC 224 (M31) and its dwarf companion NGC 221
(M32). They have accurate distances and different morphological
types, thus they are good prototypes for our investigation.

{\it NGC 221:} the observed $V,I,K,F160W$ bands SBF are well reproduced
by SSP models with an age between $5$ and $9~Gyr$ and a nearly
solar chemical content. These results are
very similar to those found by TFWG00 and TF02, i.e. $t=4.9\pm1.3 ~
Gyr$ with $[Fe/H]=-0.01\pm0.05$, and $t=3.8~ Gyr$ with $[Fe/H]=-0.04$,
respectively.

{\it NGC 224:} $V,I,F160W$ bands SBF data of the bulge of this galaxy are
consistent with SSP models of fairly old age ($t\sim15~Gyr$) and solar
metallicity. This is in agreement with Bica, Alloin \& Schmidt
(1990) and Davidge (2001), 
who find that the bulge of this galaxy is dominated by an old
population with metal content $-0.5<[Z/Z_{\sun}]<0.0$. There is a
discrepancy with TF02 age estimation ($t\sim5~Gyr$). However, Davidge
(1997) finds evidence for an age gradient in the center of NGC 224 with an
intermediate age population which is probably more concentrated in the
center with respect to the main body of the bulge; such population
should contribute only for a small fraction (10-20\%) to the total
visible flux.

It is difficult to derive information from the $K$-band SBF, due to
the already quoted overlap of the models.  However, the
$\bar{M}_K$ data are consistent with the ($t,Z$) values inferred from
$V$ and $I$ bands estimations.
Some remarkable indications on NIR
data can be found by comparing the SSP models which reproduce the
$V,I$ bands SBF with the work by Davidge (2001). Observing the
brightest AGB stars in M31 bulge, he finds that the bulk of these
stars has $K \simeq -8.4~mag$. By adopting in our SSP models
($t=15~Gyr$, $Z=Z_{\sun}$) the mass loss calibration suggested in
paragraph 4.2.2 (i.e. $\eta_{AGB}$=0.8), we find that the tip of the
AGB is as bright as $K \simeq -8.2~ mag$, a value which is in very
good agreement with the observational results by Davidge.

Among the other galaxies, here we discuss some examples which support
the consistency of our estimates in the various bands and that they
are concordant with values proposed in the literature (see Table~9).

{\it NGC 891:} it is the most metal poor galaxy in our sample. We derive
$Z\sim Z_{\sun}/2$ and age $11-13~Gyr$;

{\it NGC 1316:} the $V,I,F160W$ bands SBF magnitudes suggest a young
age ($t\sim 5~ Gyr$) and a nearly solar metallicity;

{\it NGC 3377:} from $V,I$-band SBF measurements it appears to be
consistent with a nearly solar composition and relatively young age
($\sim5~Gyr$);

{\it NGC 3379:} an age of the order of $11-13~Gyr$ and a solar
chemical composition is derived from the SBF measurements in two bands
($K,F160W$). On the other hand, $V$ and $I$-band SBF data would
require a somehow younger age ($\sim5-9~Gyr$) and a higher metallicity
($Z>Z_{\sun}$). Such discrepancy will disappear later on, using
SBF-color data;

{\it NGC 4382:} only SBF measurements in the $I$-band are available
and it seems to be located slightly out of the models grid, in the
area of the diagram where we expect to find stellar systems with
 $Z \gsim 2 Z_{\sun}$ and age younger than $5~Gyr$;

{\it NGC 4472:} $I,F160W$ bands SBF measurements are in
agreement with an SSP computed assuming $Z=Z_{\sun}$ and old age
($\sim 15~Gyr$), while a slightly higher metallicity and lower ages
are suggested by $V$ and $K$ bands.

For some galaxies the slight discrepancies in the values derived from
different bands could be related to the high uncertainty of their
distance moduli, coupled with a different sensitivity of SBF in each
band to age and metallicity. Concerning distances, we mention
also the case of the galaxy NGC 4649, which is located out of the grid
of models in the $\bar{M}_I$ panel.  As discussed by Ferrarese et
al. (2000) the NGC 4649 Virgo-subcluster is not a well defined
physical association of galaxies, thus probably the group distance
reported in Table~9 is not an appropriate choice for the actual
distance of the galaxy.  A direct measure of the distance of NGC 4649
comes from PNLF; if we adopt this value the inferred age and
metallicity are estimated to be $t \sim 5-9~Gyr$ and
$Z=2Z_{\sun}$. However, we recall that Jacoby et al. (1992) and
recently Ciardullo et al. (2002) remark that the PNLF distances seem
to be systematically shorter than SBF distances.  For this reason we
do not report the last estimations in Table~9.

We obtain a discrepancy in the age and metallicity derived from
$\bar{M}_I$ and $\bar{M}_{F160W}$ for NGC 4278 and NGC 3384. For NGC
4278 the situation is similar to that of NGC 4649; for NGC 3384 such
discrepancy could be related to the multi-component structure of this
galaxy (Busarello et al. 1996), coupled with the fact that HST SBF
measurements refer to the very inner region of the galaxy while 
an outer region is observed by T01.

The above results are summarized in Table~9 (columns 7--8).  Clearly,
the comparison with TFWG00 and TF02 evaluations should not be
over-interpreted; nonetheless the fair agreement obtained is
remarkable and supports the reliability of our models.

Although $\bar{M}_{X}$ versus $(V-I)_0$ planes can be used to set useful
(though not stringent) constraints to the evolutionary properties of
the main stellar population present in a galaxy, the drawback is
mainly the dependence on the adopted distance and on the partial
degeneration age-chemical composition founded at higher metallicity
($Z\geq 0.01$).
To solve this ambiguity {\it SBF color} versus {\it integrated color}
diagrams are useful, since they appear to be strongly dependent on
stellar populations properties (age and metallicity), and even more
remarkably both quantities do {\it not} depend on distance
determinations.

To quantify this theoretical expectation, in Fig.~\ref{fig17} we plot
the selected theoretical SBF-color (involving $VIHK$ bands) versus the
theoretical integrated color $(V-I)_0$. Theory predict up to $3.5~
mag$ difference in $(\bar{M}_V-\bar{M}_K)$ for populations with
extreme $Z$ values, while nearly $3~mag$ in the
$(\bar{M}_I-\bar{M}_K)$. Therefore, for stellar population analysis,
it would be of the greatest interest to gain SBF measurements to built
up the $(\bar{M}_V-\bar{M}_K)$ color.

In Figg.~\ref{fig18} and \ref{fig19} observational data
are compared to the theoretical SBF models in the
$(\bar{M}_V-\bar{M}_I)-(V-I)_0$ and 
$(\bar{M}_I-\bar{M}_K)-(V-I)_0$
planes.  In both figures the bulk of the galaxies is clearly
concentrated near the position of the SSP having solar chemical
composition and ages greater than $10~Gyr$, i.e. the expected locus
for early-type galaxies.

The models in the diagram of Fig.~\ref{fig18} are again sharply broken
into two regions which identify different ranges in metallicity. This
feature is substantiated by the location of the observational
data: GGCs, which occupy the low-metallicity portion, and
galaxies at redder $(V-I)_0$. As age increases, we expect a redder
$(V-I)$ and bluer $(\bar{M}_V-\bar{M}_I)$, confirming results by BVA01.

As tracer of stellar population features, $(\bar{M}_V-\bar{M}_I)$ is
not particularly accurate because of the superimposition of models
with $0.01 \leq Z\leq 0.04$ clearly seen in Fig.~\ref{fig18}.  If one
requires to disentangle the age-metallicity degeneracy, a better
choice of SBF colors should be made. For example,
$(\bar{M}_I-\bar{M}_K)$ and $(\bar{M}_V-\bar{M}_K)$ should be
preferred. Concerning the $(\bar{M}_I-\bar{M}_K)$ color,
Fig.~\ref{fig19} shows that the separation between models with
different chemical composition is high enough to allow an evaluation
of metallicity in spite of the non-negligible mean error of the
observational SBF color. In particular, by adopting this kind of
diagram the ambiguity shown in the $K$-band SBF (Fig.~\ref{fig16}) for
some galaxies can be solved. For instance, NGC 224 properties are now
derived by simple interpolation of the SBF models. Similarly to NGC
224, we can take advantage by this diagram, and the age
metallicity degeneration is fairly removed for several other
galaxies. In Table~9 (columns 9--10) we summarize the results,
reporting the age and metallicity values as inferred from the location
of the SBF measurements of each given galaxy in the
$(\bar{M}_I-\bar{M}_K)-(V-I)_0$ plane.

One may note that the fair agreement found between the results in
column (7--9) and (8--10) in Table~9 supports the general consistency
of the two SBF methods. However, few galaxies show some differences.
This is not surprising because several sources of uncertainty are at
work, in a differential way, inside the two methods. Let us recall few
of them: i) the distance moduli, ii) the presence of secondary
populations, and iii) the reddening. Each of these requires evaluations
or assumptions which affect the location of the observational points
in the two diagrams.

Before concluding this Section, let us emphasize that all our
results come from single-burst stellar population models.  On this
respect, we recall that by comparing their models with observations in
the plane $(\bar{M}_I-\bar{M}_K)$ versus $(\bar{M}_V-\bar{M}_I)$,
BVA01 conclude that composite stellar populations appear necessary for
those galaxies with $(\bar{M}_V-\bar{M}_I)\leq2.3$. On the other
hand, the present models for simple stellar populations seem to match,
at least within the observational uncertainties, the locus of the
available measurements (see Fig.~\ref{fig20}); in particular, the two
Local Group galaxies NGC 221 and 224 are well fitted. Of course, in
order to better reproduce the $SBF$ of each individual galaxy (see for
example the galaxy with the highest $\bar{M}_V-\bar{M}_I$: NGC 4472),
it could be useful to introduce composite stellar populations, even if
this implies to adopt {\it ad hoc scenarios} (see BVA01).

\section{Useful relations}

According with the above discussion, the single-burst models proposed
in this work can be reliably adopted for both distance estimations and
stellar population analysis. Some useful expressions can be given to
obtain a straightforward determination of the SBF magnitudes once the
integrated $(V-I)_0$ color of a galaxy has been measured. The least-squares
method applied to theoretical models with ages $t \geq 9~Gyr$ and
metallicity $Z\geq0.001$ (equivalently $(V-I)_0\geq0.95$) provides:

\begin{equation}
\bar{M}_{V} = (0.43\pm 0.10) + (5.26\pm 0.22)[(V-I)_0 -1.15]
\end{equation}

\begin{equation}
\bar{M}_{R} = (-0.52\pm 0.14) + (5.37\pm 0.31)[(V-I)_0 -1.15]
\end{equation}

\begin{equation}
\bar{M}_{I} = (-1.74\pm 0.23) + (3.93\pm 0.50) [(V-I)_0 -1.15]
\end{equation}

\begin{equation}
\bar{M}_{K} = (-5.22\pm 0.23) - (3.87\pm0.50) [(V-I)_0 -1.15].
\end{equation}

The relations presented here show a very good agreement for both $V$
and $I$ bands with the empirical calibrations by T01, and BVA01.  The
calibration of the $R$-band is fairly similar to the one presented by
BVA01.

Concerning the $\bar{M}_{K}$ calibration, it appears quite different
from the empirical result by LGC02.  In particular, the theoretical
calibration predicts an opposite trend with respect to the empirical
one.  This opposite behavior has been also discussed by LGC02; here we
just point out that the age of the observed stellar population may
play a relevant role.  In fact Fig.~\ref{fig15}(b) shows an age
dependence of the $K$-band SBF amplitudes much larger than those
expected in $V$ and $I$ bands (Fig.~\ref{fig15}(a), and
Fig.~\ref{fig14}). More complex models (e.g LGC02), and further
observations in the $K$-band are required to clarify this
discrepancy. A similar result is obtained for $\bar{M}_{F160W}$ when
the calibration is compared to the empirical one recently found by
Jensen et al.  (2003). Thus, the fact that NIR $SBF$ age and
metallicity variations have no degenerate effects, make quite
problematic to derive accurate one-parameter SBF calibrations for
these wavelengths.

In conclusion, the theoretical $\bar{M}_X$ versus $(V-I)_0$
relations agree well (but NIR filters) with the empirical
relations, even in the simple adopted scenario of single-burst
populations.

\section{Conclusions}

In this paper we have developed a new theoretical model for SBF
predictions of simple stellar populations (single-burst, single
chemical composition) with age ranging from 5 to 15 $Gyr$, and
metallicity from $1/200~Z_{\sun}$ to $2~Z_{\sun}$, relatively to the
standard $UBVRIJHK$ ground-based filters and the $F439W$, $F555W$,
$F814W$, $F110W$, $F160W$, and $F222M$ of the HST photometric system
({\it WFPC2} and {\it NICMOS}). The model relies on the stellar
population synthesis code developed by Brocato et al. (2000),
optimized for a fine simulation of the latest (brightest) phases of
stellar evolution, which has strong influence on SBF magnitudes.

The procedure adopted to compute the SBF amplitudes is different from
previous theoretical works, and it has the advantage of being quite
similar to the observational approach, currently used to measure the
pixel to pixel fluctuation in real galaxies.

With this tool we simulated the effect of the mass loss processes
along the RGB, the AGB and the TP. In particular, the impact of
mass loss on the prediction of SBF magnitudes has been investigated,
finding that a careful treatment of this quantity is fundamental in
determining reliable SBF evaluations.

According to formalism by Reimers, we adopted $\eta_{RGB}=0.4$ for the
RGB stars. An independent calibration of the efficiency of mass loss
along the AGB was performed by using the SBF technique.  A set of
models with different assumption on $\eta_{AGB}$ was computed.  The
observed SBF measurements of GGCs and galaxies efficiently constraint
the theoretical expectations, showing that the data are well
reproduced by SSP of the proper age and metallicity if
$\eta_{AGB}=0.8$. Such value well agree with observational data (Le
Bertre et al. 2001), and impressively reproduces the absolute $K$-band
magnitude of the stars at the tip of the AGB in NGC 224. To our knowledge
this is the first time that SBF measurements are used to constraints
stellar evolution models of AGB stars.

We also compared our models with previous works (BVA01, LCG02), finding
a substantial general agreement in spite of the different stellar
evolution models adopted in the computations of the SSPs which generate
the SBF evaluations.

The impact on SBF, due to the assumption on the IMF, is investigated
and quantified by adopting a set of reasonable values of the exponent
of a Scalo-like mass distribution. The consequences of cutting the IMF
at a low mass limit are also presented. As expected, it is found that
the contribution of upper part of the MS could not be neglected.

We investigate the effects on SBF predictions of different assumptions
on the relation to transform theoretical quantities ($log~L/L_{\sun}$
and $log~T_{eff}$) to the observed magnitudes and colors. We find that
the G87, CGK97, and LCB97 compilations provide quite similar SBF
predictions for models with $Z<0.01$, while sizeable differences
should be expected if SSP of high $Z$ are involved.

Once the theoretical framework defining the SSP models is fixed, we
successfully checked the capability of deriving distances by
our SBF predictions. This is done for a sample of GGCs for which
SBF measurements are available and with a sample of properly selected
galaxies of known distances.

As a result, a set of calibrations of absolute SBF magnitudes for
classical $UBVRIJHK$ photometric bands and for selected HST filters is
provided. The quoted test with other distance indicators (Cepheids,
Planetary Nebula Luminosity Function, SBF empirical relations, etc.) shows
that the degree of accuracy and precision of the present theoretical
calibration of the absolute SBF magnitudes is sufficient to
justify the use of these theoretical SBF as a {\it primary distance
indicator}.

On the other side, we tested the use of SBF as tracers of stellar
populations properties. In particular, we derived ages and
metallicities for a subsample of galaxies for which
SBF observations and reliable distances are available.
The results appear very promising, since the obtained values are
comparable with evaluation derived by other methods.

Moreover, {\it SBF color} versus {\it integrated color} diagrams are
proposed as particularly useful in removing the well known
age-metallicity degeneracy (e.g. Worthey 1994) which affects our
knowledge of remote stellar systems.

Finally, we suggest that future observations should take advantage of
the high sensibility of SBF measurements as stellar population tracers
in the blue side of the spectral energy distribution of galaxies. This
would be a meaningful scientific target for ground-based telescopes
with high efficiency in the $B$ wavelength range (VST) or for
telescope located out of the earth atmosphere (HST, NGST).

\acknowledgments

{\it Acknowledgments :} Financial support for this work was provided
by MIUR-Cofin 2000, under the scientific project ``Stellar Observables
of Cosmological Relevance''. This project made use of computational
resources granted by the Consorzio di Ricerca del Gran Sasso according
to the Progetto 6 'Calcolo Evoluto e sue Applicazioni (RSV6)' -
Cluster C11/B.  

We thank the anonymous referee for a careful reading
of the manuscript and for several useful suggestions for clarifying
the presentation; his/her constructive report helped us to improve the
paper.

\appendix
\section{APPENDIX: The Procedure} \label{appendix}

For sake of clearness we outline here the major steps of the procedure
we have adopted to derive SBF magnitudes.

As a first step one needs to evaluate the total flux emitted by a stellar
population generated by a single burst of star formation. In our model
we make the assumption that the integrated light is dominated by the
emission generated by stars of the simulated stellar system. This
implies that i) no source of non-thermal emission are at work, ii) the
thermal emission by interstellar gas gives no sizeable contribution to
the integrated flux, and iii) the absorption from dust and gas are
negligible. On this base, the integrated flux in a selected
photometric filter ($X$) can be obtained by knowing two quantities:

\begin{equation}
f_{X} [L(m,t,Y,Z),T_{eff}(m,t,Y,Z),Y,Z],
\end{equation}

\begin{equation}
\Phi(m,N)
\end{equation}
where the $f_{X}$ is the flux emitted by a star of mass $m$, age
$t$ (=stellar system age), luminosity $L$, effective temperature
$T_{eff}$, and chemical composition ($Y$, $Z$). Obviously, $f_{X}$ is
defined by an extended library of homogeneous and self-consistent
stellar evolutionary tracks (see Section 2), but it also depends on the
adopted transformation tables (see par. 4.3).

The second quantity $\Phi$ is strictly related to the IMF and
represents the number of stars with mass $m$ in a population globally
constituted of $N$ stars with age $t$ and chemical
composition ($Y$, $Z$).
A fundamental point is that the distribution of star
masses along the IMF is obtained by using a Monte Carlo random method.
A special attention was payed to ensure that, when a set of simulations
with the same input parameters is computed, each distribution is fully
independent from previous extractions.

Both the quantities are properly combined and integrated by the stellar
population code to derive the total integrated flux $F_{X}$ of the
simulated stellar system in the selected photometric band:

\begin{equation}
F_{X}[N,t,Y,Z] = \sum _{i=1}^N f_{X}^{i},
\end{equation}
thus this quantity evaluates absolute integrated flux of a
stellar population of $N$ stars with age $t$, chemical
composition ($Y$, $Z$), and given IMF, as located at the reference
distance of $10~pc$.

If a number $N^*_{sim}$ of simulations with exactly the same input
parameters are computed, the resulting $F_{X}^j [N,t,Y,Z]$ is {\it not}
expected to be constant since the initial mass distribution of each
simulation is randomly generated. Thus, even if we adopt the same IMF
law, each $j$-$th$ mass distribution is fully independent from
previous extractions. As a consequence, the resulting distribution of
the stars in the CMD changes. In particular, the bright stars on the
RGB/AGB (i.e. the most relevant contributors to the integrated light)
are expected to undergo to sizeable variations in number and CMD
position, and this qualitatively explains the fluctuations of $F_{X}$.

In principle, this can be considered quite similar to what happen in
observing an ideal galaxy where the same population is spreaded in a
number of pixels. Under the assumption of a ideal galaxy constituted
of a homogeneous stellar population, the number and the luminosity of
bright stars may change moving from pixel to pixel again due to
Poisson statistical fluctuations. Thus, the measured flux in each
pixel is expected to be affected by the fluctuations (TS88).

By following the formalism by Tonry and collaborators, we consider
the amplitude of the expected fluctuations (SBF) $\bar{F}_{X}$ as the
ratio of the variance to the average integrated flux, i.e.:

\begin{equation}
\bar{F}_{X} \equiv \frac{ \langle F_{X} - {\langle F_{X} \rangle}\rangle^{2}}{\langle F_{X} \rangle}= \frac{\sum_{j} [ F_{X}^{j} - \langle F_{X} \rangle ]^{2}}{N^*_{sim}-1} \cdot \frac{1}{\langle F_X \rangle},
\end{equation}
where the average flux $\langle F_{X} \rangle$ of the selected
population is given by:

\begin{equation}
\langle F_{X} \rangle=\frac{\sum_{j} F_{X}^{j}}{N^*_{sim}}.
\end{equation}

From the operational point of view, for each set of parameters
(age, chemical composition, IMF, etc.) which identifies a stellar
population model, we compute a large number of independent simulations
($N^*_{sim}=5000$) deriving the $F_{X}^j$ for each model. At this
point the SBF flux is computed directly using the right terms of
equation $(A4)$ and the definition of $(A5)$.  Finally, the SBF flux
is converted in the SBF magnitude $\bar{M}_{X}$ in usual way, being:

\begin{equation}
\bar{M}_{X} = -2.5~log \bar{F}_{X}.
\end{equation}

In order to compare our method for computing SBF with those
obtained applying the ``usual'' procedure (i.e. to integrate the
luminosity function times the squared flux per luminosity bin and
divide by the total flux: TS88, BVA01, etc.), we directly computed,
for several input parameters, the SBF values by using both our method
and the usual one.  To provide a quantitative example, we report in
Table~A1 the results for stellar population models with $t=15~ Gyr$
and $Z=Z_{\sun}$.

The procedure outlined here gives SBF amplitudes in agreement with
those obtained from the usual method, when the number of stars
($N_{star}$) in each simulation is so large that all the brightest
bins in the luminosity function have a number of stars greater than
zero.  The condition needed for a full equivalence of the two methods
is satisfied when the limit of large numbers is approached (TS88). We
find that this limit is reached when $N_{star} \gsim 10^8$ (in the
mass interval $ 0.1 \leq {\cal M}/{\cal M}_{\sun} \leq 1.5$), just
because the probability of finding very bright stars in each
simulation becomes high enough.  As $N_{star}$ decreases, the
differences between the SBF values computed by adopting the two
methods increase.

Although the procedure outlined in this work is not particularly
efficient in terms of computational time, it has some advantages.
As a first point, the present approach to derive SBF amplitudes is
more similar to the observational way of measuring SBFs in real
galaxies.
In fact, the opportunity of generating $N_{sim}$ independent
stellar populations each of them including $N_{star}$ stars
closely resembles what it is directly measured in $N_{sim}$ pixels
of a galaxy CCD image (neglecting the seeing effects), where the
single pixel contains $N_{star}$ stars.
Here, we have typically adopted $N_{star}=10^5$
which may represent the estimated number of stars per pixel for a
galaxy in the Virgo cluster observed with a CCD having 0.3"/pix
(Section 2).
For such an assumption, the limit of large numbers is not
satisfied. This implies a relatively small shift between the SBF
amplitudes obtained with our procedure and the usual one as
shown in Table~A1 (Fig.~\ref{fig21}).

As a new interesting results we suggest that the SBF magnitudes
appear to have a small but not negligible dependence on the
population (pixel) star density. Although this dependence is only
a second order correction, it should be further investigated, both
on theoretical and observational side, for a deeper understanding
and a refining of the SBF method. In particular, we suggest that
the present approach would be useful when dealing with SBF
measurements obtained by pixels where the expected number of stars
is relatively low (nearby galaxies, high resolution telescopes,
loose galaxies) when compared with typical observations in
ellipticals.

As a final consideration on the capability of the present way of
computing SBF we note that it is also particularly efficient in
handling the details of all the stellar evolutionary phases.  In fact,
each independent stellar population is computed by leaving to the
statistics the possibility of distributing stars in the CMDs,
according to their proper evolutionary pattern. On the other hand, the
usual way makes use of predefinite isochrones which fix, by
definition, one single pattern in the CMD (i.e. a fixed luminosity
function) for each set of input parameters ($t,Z$). A relevant example
is the HB stars distribution. As previously shown by Rood et al. (1973), the
HB morphology can hardly be described by uniparametric (mass) curves
as assumed when dealing with isochrones.  In our procedure we always
compute a synthetic HB which reproduces the color spread and relative
star counts observed in real HBs (Brocato et al. 2000).  Moreover,
isochrones are generally computed by assuming a fixed value of mass
loss rate along the RGB.  On the contrary, in the present stellar
population synthesis code the mass loss and its spread around an
average value can be chosen as free input parameters (section 4.2.1).

\newpage


\begin{figure}
\caption{
Simulated Color-Magnitude Diagrams for a fixed age (t=15 $Gyr$) and for
three values of metallicity: $Z$=0.001 (triangles), 0.01 (stars),
0.02 (circles). A solar metallicity population with $t=15~Gyr$, 
transformation tables are from Castelli, Gratton \& Kurucz (1997).
\label{fig1} }
\end{figure}

\begin{figure}
\caption{
SBF amplitude $\bar{M}_{I}$ as a function of the number
of Monte Carlo cycles. For N$_{sim}>$1000, $\bar{M}_{I}$ varies around the
asymptotic value (the permitted region is delimited by the dashed lines)
by an amount well within the typical observational uncertainties ($\sim$0.05).
The arrow indicates the value N$_{sim}^{*}$=5000 adopted in our calculations
(see text).}
\label{fig2}
\end{figure}

\begin{figure}
\caption{
{\it (a)} Predicted SBF absolute magnitudes within $UBVRIJHK$
ground-based filters for reference models aged $15~Gyr$ (solid line) and
$5~Gyr$ (dashed line) versus the metal content.
LCB97 transformations were adopted.
The selected passbands are marked with different symbols, as labeled.}
{\bf {\it (b)} As in panel {\it (a)} but for HST filters. In this case
we adopt OL00 transformations.}
\label{fig3}
\end{figure}

\begin{figure}
\caption{
Predicted SBF absolute magnitudes within $UBVRIJHK$ filters for
reference models versus the age of the stellar population ($Z=0.02$
solid lines, $Z=0.0001$ dashed lines).
The selected passbands are marked with different symbols, as labeled.
\label{fig4}}
\end{figure}

\begin{figure}
\caption{
Comparison between SBF synthetic models with similar age: LGC02 models with
$t=17~Gyr$ (dotted line and four pointed stars);
BVA01 models $t=15.8~Gyr$ (dashed line and three pointed stars);
present work models with $t=15~Gyr$ (solid line  and filled circles).
\label{fig5}}
\end{figure}

\begin{figure}
\caption{
The effect of the mass loss variations along the RGB on $UVIK$ bands SBF
magnitudes. The mass loss is parameterized by the $\eta$ coefficient
of the Reimers formula. The dotted line refers to $Z=0.04$, the solid line
to $Z=0.02$, and the dashed line to $Z=0.01$.
All models refer to $t=15~Gyr$ populations.
For graphical reasons we shifted all non solar models to match the SBF values 
of the solar model with $\eta$=0.2.
\label{fig6}}
\end{figure}

\begin{figure}
\caption{
The behavior of SBF magnitudes in the $IJHK$ bands as function of
the $(V-I)_0$ integrated color. A stellar population of solar chemical composition
is assumed.
The size of the symbols marks increasing values of $\eta_{AGB}$ (0.4,
0.8, 1.6, and 3.2); see also the orientation of the labeled arrow. The
other arrow indicates the orientation of increasing ages (5, 9, 11, 13,
15 $Gyr$) that are also linked by dotted lines. 
\label{fig7}}
\end{figure}

\begin{figure}
\caption{
Theoretical models (open symbols) obtained by adopting
the labeled values of $\eta_{AGB}$ are compared with the
empirical calibration by T01 (slanted solid line). The
horizontal solid line refers to the observational locus of GGCs. Each symbol
indicates a different chemical composition: $Z=0.0001$ (three-pointed
stars); $Z=0.001$ (four-pointed stars); $Z=0.01$ (five-pointed stars);
$Z=0.02$ (triangles); $Z=0.04$ (squares). 
The symbol size increases according to age ($t=5$, 9, 11, 13, and $15~Gyr$). The arrows
indicate the directions of the evolution with age and metallicity.
Both the horizontal and the slanted lines suffer of an
uncertainty in the zero point as large as $0.15$ and $0.20~mag$, respectively
(shadowed regions).
\label{fig8}}
\end{figure}

\begin{figure}
\caption{
$UBVI$ bands SBF magnitudes versus the number density of PAGB stars relative to
HB stars ($N_{PAGB}/N_{HB}$). Solid lines refer to the BS97 and dashed
lines to VW94 PAGB models. The shadowed region indicates the proper
ratio predicted for our reference models.
A solar metallicity population ($Z=0.02$)
with age $t=15~ Gyr$ is assumed.
\label{fig9}}
\end{figure}

\begin{figure}
\caption{
The effects of transformations tables on SBF amplitudes
are illustrated as a function of metallicity.
The residuals are computed as: $\Delta \bar{M}=SBF_X-SBF_{LCB97}$,
where $X=CGK97$
(dotted line, filled triangles) or $X=G87$ (long-dashed line, open
squares). An age of $t=15~Gyr$ is considered.
\label{fig10}}
\end{figure}

\begin{figure}
\caption{ The solar metallicity isochrone (t=15 $Gyr$)
is transformed with LCB97 (bold solid line), CGK97 (dotted line), G87
(dashed line), and Alonso et al. 1996-1999 (long-dashed line). The three
panels show respectively: $V$ versus $(V-I)_0$, $I$ versus $(V-I)_0$, and $K$
versus $(V-I)_0$. $V,I$, and $K$ are the absolute magnitudes.  The star
indicates where the extrapolation on colors for CGK97 tables
($T_{eff}<3500~K$) begins. Alonso et al. (1999) models have nearly the
same limit, being the minimum temperature tabulated $T_{min}=3750~K$.
\label{fig11}}
\end{figure}

\begin{figure}
\caption{
Our SBF predictions (open triangles) in $V$ and $I$ bands plotted
against the observational data of 19 GGCs (AT94, open
circles). Models of increasing age are marked with larger symbols
($t=5$, 9, 11, 13, $15~Gyr$); the
solid (long-dashed) line connects models with $t=15~Gyr$ ($t=5~Gyr$)
for different chemical compositions.  Dotted lines represent the
best-fit as derived by BVA01 (see text).
\label{fig12}}
\end{figure}
\clearpage

\begin{figure}
\caption{
Theoretical $I$-band SBF are compared to observed
data. $\bar{M}_{I}$ for observed galaxies is derived by adopting the
distance modulus from the Period-Luminosity relation of
Cepheids variables (Ferrarese et al. 2000). Symbols are as in
Fig.~\ref{fig8}.
\label{fig13}}
\end{figure}

\begin{figure}
\caption{
As in Fig.~\ref{fig13}, but for a sample of galaxies
with distances measured from other distance indicators.
Distance moduli are again from Ferrarese et al.
(2000).
\label{fig14}}
\end{figure}

\begin{figure}
\caption{
{\it (a)} As in Fig.~\ref{fig14}, but for $\bar{M}_{V}$. Galaxies with
Cepheids-based distances are added (circled dots).}
\figurenum{15}
\label{fig15}
\end{figure}

\begin{figure}
\caption{
{\it (b)} As in {\it (a)}, but for $\bar{M}_{K}$.  Filled circles are
$K'$ measurements, while open circles $K_{s}$. Galaxies with multiple
observations are still plotted using different symbols.}
\figurenum{15}
\end{figure}

\begin{figure}
\caption{
{\it (c)} As in {\it (a)}, but for $\bar{M}_{F160W}$. 
We note that NGC 4278 has distance measurements 
with a large spread, if compared to internal uncertainties 
(Ferrarese et al. 2000).}
\figurenum{15}
\end{figure}

\begin{figure}
\figurenum{16}
\caption{The subsample of galaxies with age and metallicity
estimations from the literature are plotted against theoretical models
for all ages and metallicity higher than half-solar. 
The galaxies are identified
at the top of each panel according to the increasing
integrated $(V-I)_0$ color. For each galaxy we also report 
the size of the depth of its group according to Ferrarese et al. (2000).
\label{fig16}}
\end{figure}

\begin{figure}
\figurenum{17}
\caption{Theoretical {\it SBF color} versus {\it integrated $(V-I)_0$ color}
diagrams obtained with our fluctuations models. 
Note the major excursion of the $(\bar{M}_V-\bar{M}_K)$ color
for extreme $Z$ values, compared to other SBF colors. Symbols are as
in Fig.~\ref{fig8}.
\label{fig17}}
\end{figure}

\begin{figure}
\figurenum{18}
\caption{The $(\bar{M}_V-\bar{M}_I)$ color is plotted against
the integrated color $(V-I)_0$.
Open circles refer to GGCs data, and filled circles to
galaxies. The typical error bars of GGCs data are also plotted.
\label{fig18}}
\end{figure}

\begin{figure}
\figurenum{19}
\caption{The $(\bar{M}_I-\bar{M}_K)$ color is plotted against
the integrated color $(V-I)_0$. For clarity we only plot models
with the highest metallicity considered.
\label{fig19}}
\end{figure}
\clearpage

\begin{figure}
\figurenum{20}
\caption{The $(\bar{M}_V-\bar{M}_I)$ versus the $(\bar{M}_I-\bar{M}_K)$ color
are plotted for the new models (symbols as in Fig.~\ref{fig8}) and observational 
data. The observed SBF amplitudes of galaxies are: $V$-band from BVA01 
(filled circles), TAL90 (open circles), and AT94 (NGC 224);
$I$-band from T01; 
$K$-band from Luppino \& Tonry (1993) and Jensen et al. (1998).
\label{fig20}}
\end{figure}

\begin{figure}
\figurenum{21}
\caption{SBF amplitudes obtained with the ``usual'' (dashed line) 
and the ``present work'' (filled circles) procedures versus $N_{star}$.}
\label{fig21}
\end{figure}
\clearpage

\pagestyle{empty}
\thispagestyle{empty}
\begin{deluxetable}{llcccccccccc}
\tablecaption{Surface Brightness Fluctuations for the reference models: ground-based filters}
\tablewidth{420pt}
\tabletypesize{\small}
\setlength{\tabcolsep}{0.07in}
\tablenum{1a}
\tablehead{
\colhead{$Z$}                & \colhead{$Y$}              & \colhead{$Age$}                &
\colhead{$\bar{M}_{U}$ }     & \colhead{$\bar{M}_{B}$ }   & \colhead{$\bar{M}_{V}$ }       &
\colhead{$\bar{M}_{R}$ }     & \colhead{$\bar{M}_{I}$ }   & \colhead{$\bar{M}_{J}$ }       &
\colhead{$\bar{M}_{H}$ }     & \colhead{$\bar{M}_{K}$ }   & \colhead{$(V-I)$ } }
\startdata
0.0001 & 0.23 & 5  & 1.711  & 0.934  &  -0.776  &  -1.771  & -2.690  &   -3.719  &   -4.456  &   -4.658  &  0.776  \\
       &      & 9  & 1.527  & 0.759  &  -0.815  &  -1.738  & -2.592  &   -3.564  &   -4.266  &   -4.457  &  0.855  \\
       &      & 11 & 1.552  & 0.774  &  -0.719  &  -1.596  & -2.409  &   -3.355  &   -4.035  &   -4.218  &  0.858  \\
       &      & 13 & 1.577  & 0.827  &  -0.636  &  -1.490  & -2.283  &   -3.216  &   -3.884  &   -4.062  &  0.871  \\
       &      & 15 & 1.534  & 0.811  &  -0.610  &  -1.425  & -2.185  &   -3.102  &   -3.752  &   -3.923  &  0.891  \\
\tableline
0.001  & 0.23 & 5  & 2.426  & 1.176  &  -0.685  &  -1.776  & -2.817  &   -4.009  &   -4.949  &   -5.184  &   0.913 \\
       &      & 9  & 2.304  & 1.131  &  -0.571  &  -1.548  & -2.464  &   -3.564  &   -4.441  &   -4.650  &   0.965  \\
       &      & 11 & 2.367  & 1.198  &  -0.474  &  -1.424  & -2.306  &   -3.375  &   -4.235  &   -4.434  &   0.977  \\
       &      & 13 & 2.183  & 1.141  &  -0.478  &  -1.404  & -2.260  &   -3.312  &   -4.158  &   -4.349  &   0.980  \\
       &      & 15 & 2.008  & 1.080  &  -0.425  &  -1.331  & -2.174  &   -3.215  &   -4.057  &   -4.244  &   0.959  \\
\tableline
0.01   & 0.255 & 5  & 3.476  & 1.801  &   0.055  &  -1.037  & -2.495  &   -4.348  &   -5.488  &   -5.839  &   1.075 \\
       &      & 9  & 3.386  & 1.875  &   0.237  &  -0.797  & -2.164  &   -4.027  &   -5.153  &   -5.542  &   1.114 \\
       &      & 11 & 3.388  & 1.887  &   0.256  &  -0.775  & -2.126  &   -3.868  &   -4.982  &   -5.339  &   1.130 \\
       &      & 13 & 3.351  & 1.902  &   0.283  &  -0.758  & -2.104  &   -3.728  &   -4.816  &   -5.152  &   1.151 \\
       &      & 15 & 3.324  & 1.927  &   0.332  &  -0.697  & -2.011  &   -3.644  &   -4.732  &   -5.082  &   1.167 \\
\tableline
0.02   & 0.289 & 5 & 4.124  & 2.247  &   0.528  &  -0.489  & -1.918  &   -4.518  &   -5.656  &   -6.145  &   1.112 \\
       &       & 9 & 4.230  & 2.352  &   0.702  &  -0.270  & -1.671  &   -4.064  &   -5.190  &   -5.633  &   1.174 \\
       &       & 11& 4.169  & 2.325  &   0.710  &  -0.238  & -1.612  &   -4.025  &   -5.150  &   -5.600  &   1.196 \\
       &       & 13& 4.111  & 2.344  &   0.759  &  -0.173  & -1.526  &   -3.883  &   -5.010  &   -5.447  &   1.211   \\
       &       & 15& 4.085  & 2.353  &   0.781  &  -0.152  & -1.513  &   -3.846  &   -4.974  &   -5.402  &   1.229   \\
\tableline
0.04    & 0.34  & 5 & 4.796  & 2.715  &   0.987  &   0.013  & -1.411  &   -4.554  &   -5.654  &   -6.203  &   1.188 \\
        &       & 9 & 4.792  & 2.763  &   1.116  &   0.193  & -1.121  &   -4.120  &   -5.241  &   -5.770  &   1.249 \\
        &       & 11& 4.797  & 2.769  &   1.140  &   0.230  & -1.045  &   -3.984  &   -5.111  &   -5.630  &   1.271 \\
        &       & 13& 4.779  & 2.804  &   1.198  &   0.299  & -0.980  &   -3.948  &   -5.083  &   -5.600  &   1.288 \\
        &       & 15& 4.740  & 2.786  &   1.198  &   0.306  & -0.960  &   -3.882  &   -5.012  &   -5.522  &   1.304 \\
\enddata
\end{deluxetable}
\clearpage

\thispagestyle{empty}
\begin{deluxetable}{llcccccccc}
\tablecaption{Surface Brightness Fluctuations for the reference models: HST filters}
\tablewidth{420pt}
\tabletypesize{\small}
\setlength{\tabcolsep}{0.05in}
\tablenum{1b}
\tablehead{
\colhead{$Z$}                & \colhead{$Y$}              & \colhead{$Age$}                &
\colhead{$\bar{M}_{F439W}$}  & \colhead{$\bar{M}_{F555W}$}& \colhead{$\bar{M}_{F814W}$}    &
\colhead{$\bar{M}_{F110W}$}  & \colhead{$\bar{M}_{F160W}$}& \colhead{$\bar{M}_{F222M}$}    &
\colhead{$(V-I)$ } }
\startdata
0.0001 & 0.23 & 5 & 1.141 & -0.794 & -2.752 & -3.442 & -4.414 & -4.566 & 0.776 \\
       &      & 9 & 0.936 & -0.834 & -2.654 & -3.302 & -4.228 & -4.370 & 0.855 \\
       &      & 11& 0.933 & -0.735 & -2.471 & -3.099 & -3.999 & -4.133 & 0.858 \\
       &      & 13& 0.981 & -0.650 & -2.345 & -2.963 & -3.849 & -3.978 & 0.871 \\
       &      & 15& 0.965 & -0.623 & -2.247 & -2.853 & -3.720 & -3.842 & 0.891 \\
\tableline
0.001  & 0.23 & 5 & 1.410 & -0.696 & -2.897 & -3.720 & -4.873 & -5.088 & 0.913 \\
       &      & 9 & 1.339 & -0.580 & -2.541 & -3.287 & -4.374 & -4.560 & 0.965 \\
       &      & 11& 1.402 & -0.482 & -2.382 & -3.103 & -4.170 & -4.346 & 0.977 \\
       &      & 13& 1.331 & -0.486 & -2.336 & -3.041 & -4.095 & -4.263 & 0.980 \\
       &      & 15& 1.233 & -0.433 & -2.249 & -2.946 & -3.995 & -4.159 & 0.959 \\
\tableline
0.01   & 0.255& 5 & 2.095 &  0.103 & -2.625 & -4.073 & -5.394 & -5.788 & 1.075 \\
       &      & 9 & 2.117 &  0.240 & -2.329 & -3.738 & -5.055 & -5.455 & 1.114 \\
       &      & 11& 2.136 &  0.272 & -2.272 & -3.596 & -4.882 & -5.252 & 1.130 \\
       &      & 13& 2.161 &  0.327 & -2.192 & -3.472 & -4.747 & -5.097 & 1.151 \\
       &      & 15& 2.176 &  0.366 & -2.120 & -3.381 & -4.644 & -4.994 & 1.167 \\
\tableline
0.02   & 0.289 & 5& 2.466 &  0.496 & -2.246 & -4.583 & -5.605 & -6.024 & 1.112 \\
       &       & 9& 2.567 &  0.675 & -1.937 & -3.737 & -5.120 & -5.515 & 1.174 \\
       &       &11& 2.534 &  0.682 & -1.890 & -3.700 & -5.081 & -5.478 & 1.196 \\
       &       &13& 2.547 &  0.733 & -1.789 & -3.564 & -4.936 & -5.327 & 1.211 \\
       &       &15& 2.555 &  0.753 & -1.770 & -3.531 & -4.897 & -5.285 & 1.229 \\
\tableline
0.04   & 0.34 & 5 & 2.890 &  0.944 & -1.791 & -4.202 & -5.741 & -6.251 & 1.188   \\
       &      & 9 & 2.930 &  1.077 & -1.516 & -3.779 & -5.272 & -5.756 & 1.249 \\
       &      & 11& 2.935 &  1.104 & -1.437 & -3.647 & -5.120 & -5.591 & 1.271 \\
       &      & 13& 2.967 &  1.163 & -1.395 & -3.613 & -5.079 & -5.545 & 1.288\\
       &      & 15& 2.947 &  1.163 & -1.368 & -3.550 & -5.005 & -5.465 & 1.304\\
\enddata
\end{deluxetable}
\clearpage

\begin{deluxetable}{ccccccccc}
\tabletypesize{\small}
\tablenum{2}
\tablewidth{350pt}
\tablecaption{
SBF amplitudes for different assumptions on the lowest mass limit
(${\cal M}_{low}$) for a population model with $t=15~ Gyr$, $Z=0.02$,  $Y=0.289$}
\tablehead{\colhead{${\cal M}_{low}$} & \colhead{$\bar{M}_{U}$} &
\colhead{$\bar{M}_{B}$}              & \colhead{$\bar{M}_{V}$} &
\colhead{$\bar{M}_{R}$}              & \colhead{$\bar{M}_{I}$} &
\colhead{$\bar{M}_{J}$}              & \colhead{$\bar{M}_{H}$} &
\colhead{$\bar{M}_{K}$}              }
\startdata
0.85 &  3.886 & 2.146 & 0.571 & -0.374 & -1.716 & -3.914 & -5.024 & -5.425    \\
0.60 &  4.106 & 2.361 & 0.775 & -0.174 & -1.534 & -3.769 & -4.896 & -5.306    \\
0.40 &  4.117 & 2.377 & 0.798 & -0.142 & -1.487 & -3.717 & -4.845 & -5.255    \\
0.10 &  4.112 & 2.385 & 0.811 & -0.122 & -1.451 & -3.668 & -4.798 & -5.205    \\
\enddata
\end{deluxetable}
\clearpage

\begin{deluxetable}{ccccccccc}
\tabletypesize{\small}
\tablenum{3}
\tablewidth{320pt}
\tablecaption{
SBF amplitudes for different IMF slopes $x$ (see text)
 for a population model with $t=15~ Gyr$, $Z=0.02$, $Y=0.289$}
\tablehead{\colhead{$x$} &
\colhead{$\bar{M}_{U}$} & \colhead{$\bar{M}_{B}$} &
\colhead{$\bar{M}_{V}$} & \colhead{$\bar{M}_{R}$} &
\colhead{$\bar{M}_{I}$} & \colhead{$\bar{M}_{J}$} &
\colhead{$\bar{M}_{H}$} & \colhead{$\bar{M}_{K}$} }
\startdata
0.35 &  4.029 &  2.300 &   0.733 &  -0.198 &  -1.552 &  -3.889 &  -5.011 &  -5.441 \\
1.35 &  4.085 &  2.353 &   0.781 &  -0.153 &  -1.513 &  -3.846 &  -4.974 &  -5.402 \\
2.35 &  4.170 &  2.441 &   0.873 &  -0.049 &  -1.385 &  -3.703 &  -4.838 &  -5.263 \\
\enddata
\end{deluxetable}

\clearpage

\begin{deluxetable}{ccccccccccc}
\tablewidth{370pt}
\tabletypesize{\small}
\setlength{\tabcolsep}{0.07in}
\tablenum{4}
\tablecaption{
SBF amplitudes for simple stellar populations with
age $t=15~ Gyr$ and different values of Reimers' coefficient $\eta$}
\tablehead{\colhead{$Z$} & \colhead{$\eta$} &
\colhead{$\bar{M}_{U}$} & \colhead{$\bar{M}_{B}$} &
\colhead{$\bar{M}_{V}$} & \colhead{$\bar{M}_{R}$} &
\colhead{$\bar{M}_{I}$} & \colhead{$\bar{M}_{J}$} &
\colhead{$\bar{M}_{H}$} & \colhead{$\bar{M}_{K}$} &\colhead{$(V-I)$} }
\startdata
0.01 &  0.2 &   3.342 &    1.915 & 0.329  & -0.682 & -2.054 &  -3.853 & -4.976 & -5.341  & 1.173 \\
     &  0.4 &   3.324 &    1.927 & 0.332  & -0.697 & -2.011 &  -3.644 & -4.732 & -5.082  & 1.167 \\
     &  0.6 &   3.009 &    1.885 & 0.367  & -0.623 & -1.890 &  -3.542 & -4.634 & -4.990  & 1.148 \\
     &  0.8 &   2.284 &    1.620 & 0.417  & -0.540 & -1.850 &  -3.563 & -4.673 & -5.036  & 1.078 \\
\tableline
0.02 &  0.2 &   4.092 &    2.345 & 0.774 &  -0.159 & -1.552 &  -4.035 & -5.163 & -5.618 &  1.234 \\
     &  0.4 &   4.085 &    2.353 & 0.781 &  -0.152 & -1.513 &  -3.846 & -4.974 & -5.402 &  1.229 \\
     &  0.6 &   4.023 &    2.375 & 0.818 &  -0.116 & -1.459 &  -3.699 & -4.827 & -5.237 &  1.221 \\
     &  0.8 &   3.578 &    2.273 & 0.791 &  -0.122 & -1.418 &  -3.621 & -4.756 & -5.164 &  1.201 \\
\tableline
0.04 &  0.2 &   4.746 &    2.780 & 1.186  &  0.288 & -1.004 &  -4.062 & -5.192 & -5.719 &  1.311 \\
     &  0.4 &   4.740 &    2.786 & 1.198  &  0.306 & -0.960 &  -3.882 & -5.012 & -5.522 &  1.304\\
     &  0.6 &   4.646 &    2.758 & 1.188  &  0.305 & -0.945 &  -3.761 & -4.889 & -5.388 &  1.292\\
     &  0.8 &   4.186 &    2.638 & 1.160  &  0.307 & -0.898 &  -3.696 & -4.835 & -5.338 &  1.260\\
\enddata
\end{deluxetable}

\clearpage

\begin{deluxetable}{lcccccccccl}
\tablewidth{400pt}
\tabletypesize{\small}
\setlength{\tabcolsep}{0.07in}
\tablenum{5}
\tablecaption{
SBF amplitudes for simple stellar populations with
$Z=0.02$, $Y=0.289$, and different assumptions for
$\eta_{AGB}$}
\tablehead{\colhead{$\eta_{AGB}$} & \colhead{$Age$} &
\colhead{$\bar{M}_{U}$} & \colhead{$\bar{M}_{B}$} &
\colhead{$\bar{M}_{V}$} & \colhead{$\bar{M}_{R}$} &
\colhead{$\bar{M}_{I}$} & \colhead{$\bar{M}_{J}$} &
\colhead{$\bar{M}_{H}$} & \colhead{$\bar{M}_{K}$} &\colhead{$(V-I)$} }
\startdata
0.4 & 5 & 4.122 &   2.245 &  0.526 &  -0.499 & -2.047 & -5.145 &  -6.275 &   -6.803 &   1.137 \\
    & 9 & 4.226 &   2.350 &  0.700 &  -0.279 & -1.778 & -4.607 &  -5.731 &   -6.235 &   1.192   \\
    &11 & 4.168 &   2.325 &  0.711 &  -0.241 & -1.684 & -4.448 &  -5.575 &   -6.075 &   1.208   \\
    &13 & 4.111 &   2.344 &  0.760 &  -0.175 & -1.598 & -4.325 &  -5.455 &   -5.950 &   1.222   \\
    &15 & 4.084 &   2.353 &  0.782 &  -0.153 & -1.550 & -4.106 &  -5.236 &   -5.704 &   1.237   \\
\tableline
0.8 & 5 & 4.124 &   2.247 &  0.528  & -0.489 & -1.918  &-4.518 &  -5.656 &   -6.145  &  1.112 \\
    & 9 & 4.230 &   2.352 &  0.702  & -0.270 & -1.671  &-4.064 &  -5.190 &   -5.633  &  1.174   \\
    &11 & 4.169 &   2.325 &  0.710  & -0.238 & -1.612  &-4.025 &  -5.150 &   -5.600  &  1.196 \\
    &13 & 4.111 &   2.344 &  0.759  & -0.173 & -1.526  &-3.883 &  -5.010 &   -5.447  &  1.211   \\
    &15 & 4.085 &   2.353 &  0.781  & -0.152 & -1.513  &-3.846 &  -4.974 &   -5.402  &  1.229   \\
\tableline
1.6 & 5 & 4.127 &   2.249 &  0.530 &  -0.482 & -1.849 & -4.129 &  -5.257 &   -5.697 &   1.102   \\
    & 9 & 4.232 &   2.353 &  0.702 &  -0.268 & -1.627 & -3.826 &  -4.949 &   -5.356 &   1.166   \\
    &11 & 4.173 &   2.327 &  0.712 &  -0.234 & -1.567 & -3.745 &  -4.867 &   -5.268 &   1.188   \\
    &13 & 4.116 &   2.346 &  0.760 &  -0.170 & -1.485 & -3.660 &  -4.783 &   -5.184 &   1.204    \\
    &15 & 4.088 &   2.354 &  0.781 &  -0.150 & -1.474 & -3.670 &  -4.795 &   -5.194 &   1.222   \\
\tableline
3.2 & 5 & 4.130 &  2.251 &  0.531 &  -0.478 & -1.794 &  -3.742 & -4.849 &   -5.204 &   1.093   \\
    & 9 & 4.233 &  2.352 &  0.699 &  -0.269 & -1.584 &  -3.614 & -4.731 &   -5.102 &   1.158   \\
    & 11& 4.177 &  2.330 &  0.714 &  -0.226 & -1.567 &  -3.583 & -4.703 &   -5.089 &   1.178   \\
    & 13& 4.121 &  2.352 &  0.770 &  -0.150 & -1.415 &  -3.552 & -4.681 &   -5.082 &   1.193   \\
    & 15& 4.099 &  2.377 &  0.819 &  -0.093 & -1.383 &  -3.606 & -4.741 &   -5.154 &   1.208   \\
\enddata
\end{deluxetable}
\clearpage

\begin{deluxetable}{lcccccccc}
\tabletypesize{\small}
\tablenum{6}
\tablewidth{340pt}
\tablecaption{
SBF amplitudes for different color-temperature relations (see text). An age $t=15~Gyr$ is assumed}
\tablehead{\colhead{$Z$} &
\colhead{$\bar{M}_{U}$} & \colhead{$\bar{M}_{B}$} &
\colhead{$\bar{M}_{V}$} & \colhead{$\bar{M}_{R}$} &
\colhead{$\bar{M}_{I}$} & \colhead{$\bar{M}_{J}$} &
\colhead{$\bar{M}_{H}$} & \colhead{$\bar{M}_{K}$} }
\startdata
\multicolumn {9} {l}  {LCB97 Atmospheres transformations} \\
\tableline
0.0001 & 1.533 &   0.811 & -0.610 &  -1.424 &  -2.184 &  -3.101 &  -3.751 & -3.923 \\
0.001  & 2.008 &   1.080 & -0.424 &  -1.330 &  -2.173 &  -3.215 &  -4.056 & -4.244 \\
0.01   & 3.324 &   1.926 &  0.331 &  -1.697 &  -2.011 &  -3.644 &  -4.732 & -5.081 \\
0.02   & 4.085 &   2.352 &  0.781 &  -1.152 &  -1.512 &  -3.846 &  -4.973 & -5.402 \\
0.04   & 4.740 &   2.786 &  1.198 &   0.305 &  -0.960 &  -3.881 &  -5.012 & -5.521 \\
\tableline
\multicolumn {9} {l}  {CGK97 Atmospheres transformations} \\
\tableline
0.0001 & 1.579 & 0.817 & -0.522 &  -1.271 & -2.004 & -3.142 &  -3.869 &  -4.048  \\
0.001  & 2.108 & 1.072 & -0.427 &  -1.259 & -2.050 & -3.281 &  -4.161 &  -4.368 \\
0.01   & 3.433 & 1.863 &  0.128 &  -0.850 & -1.933 & -3.661 &  -4.830 &  -5.145 \\
0.02   & 4.094 & 2.253 &  0.496 &  -0.510 & -1.474 & -3.815 &  -5.094 &  -5.416 \\
0.04   & 4.844 & 2.660 &  0.878 &  -0.125 & -1.121 & -3.803 &  -5.218 &  -5.585\\
\tableline
\multicolumn {9} {l}  {G87 Atmospheres transformations} \\
\tableline
0.0001 & 1.587 & 0.828 & -0.432 &  -1.248 &  -1.943 & \nodata & \nodata & \nodata \\
0.001  & 2.067 & 1.037 & -0.426 &  -1.274 &  -2.044 & \nodata & \nodata & \nodata \\
0.01   & 3.013 & 1.769 &  0.151 &  -0.850 &  -2.087 & \nodata & \nodata & \nodata \\
0.02   & 3.355 & 2.049 &  0.515 &  -0.468 &  -2.033 & \nodata & \nodata & \nodata \\
0.04   & 3.743 & 2.338 &  0.863 &  -0.075 &  -2.084 & \nodata & \nodata & \nodata \\
\enddata
\end{deluxetable}
\clearpage

\begin{deluxetable}{lcccclcc}
\tabletypesize{\small}
\tablenum{7}
\tablecaption{Measured SBF amplitudes $\bar{m}_V$ and  $\bar{m}_I$, and
$(V-I)_0$ color are from AT94. Other GGCs properties are from Harris (1999)}
\tablewidth{450pt}
\tablehead{
\colhead{GGC Name} & \colhead{$\bar{m}_V$} & \colhead{$\bar{m}_I$} &
\colhead{$E_{B-V}$} & \colhead{$(V-I)_0$} &\colhead{$M_{V}^{tot}$} &  \colhead{$[Fe/H]$} & \colhead{$DM$}}
 \startdata
NGC 104 (47Tuc) &13.15 $\pm$ 0.08 &   11.10 $\pm$  0.16 &    0.04 &  1.06 & $-$9.42  &  $-$0.76 &   13.37  \\
NGC 288         &14.53 $\pm$ 0.34 &   12.57 $\pm$  0.45 &    0.03 &       0.92 & $-$6.60  &  $-$1.24 &   14.69  \\
NGC 2419    &19.29 $\pm$ 0.09 &   17.72 $\pm$  0.12 &    0.11 &  0.85 & $-$9.58  &  $-$2.34 &   19.97  \\
NGC 5139        &13.67 $\pm$ 0.10 &   12.01 $\pm$  0.12 &    0.12 &  0.95 & $-$10.29 &  $-$1.62 &   13.97  \\
NGC 5272 (M3)   &14.37 $\pm$ 0.14 &   12.59 $\pm$  0.17 &    0.01 &  0.88 & $-$8.93  &  $-$1.57 &   15.12  \\
NGC 5904 (M5)   &14.04 $\pm$ 0.16 &   12.30 $\pm$  0.19 &    0.03 &  0.89 & $-$8.81  &  $-$1.29 &   14.46  \\
NGC 6093 (M80)  &14.84 $\pm$ 0.18 &   12.90 $\pm$  0.21 &    0.18 &  0.99 & $-$8.23  &  $-$1.75 &   15.56  \\
NGC 6205 (M13)  &13.68 $\pm$ 0.16 &   12.05 $\pm$  0.20 &    0.02 &  0.88 & $-$8.70  &  $-$1.54 &   14.48  \\
NGC 6341 (M92)  &13.85 $\pm$ 0.17 &   12.23 $\pm$  0.20 &    0.02 &  0.86 & $-$8.20  &  $-$2.29 &   14.64  \\
NGC 6624    &15.45 $\pm$ 0.20 &   12.40 $\pm$  0.27 &    0.28 &  1.20 & $-$7.50  &  $-$0.42 &   15.37  \\
NGC 6626 (M28)  &14.41 $\pm$ 0.18 &   11.86 $\pm$  0.22 &    0.43 &  1.01 & $-$8.33  &  $-$1.45 &   15.12  \\
NGC 6637 (M69)  &15.30 $\pm$ 0.11 &   12.91 $\pm$  0.19 &    0.16 &  1.07 & $-$7.52  &  $-$0.71 &   15.16  \\
NGC 6652        &15.43 $\pm$ 0.13 &   13.48 $\pm$  0.22 &    0.09 &  1.03 & $-$7.73  &  $-$0.96 &   15.19  \\
NGC 6656 (M22)  &12.96 $\pm$ 0.21 &   10.83 $\pm$  0.26 &    0.34 &  1.00 & $-$8.50  &  $-$1.64 &   13.60  \\
NGC 6715 (M54)  &16.93 $\pm$ 0.12 &   15.16 $\pm$  0.13 &    0.14 &  0.90 & $-$10.01 &  $-$1.59 &   17.61  \\
NGC 6723        &14.88 $\pm$ 0.13 &   13.23 $\pm$  0.22 &    0.05 &  1.01 & $-$7.86  &  $-$1.12 &   14.87  \\
NGC 7006    &17.53 $\pm$ 0.18 &   16.03 $\pm$  0.21 &    0.05 &  0.92 & $-$7.68  &  $-$1.63 &   18.24  \\
NGC 7078 (M15)  &14.65 $\pm$ 0.11 &   13.06 $\pm$  0.13 &    0.10 &  0.83 & $-$9.17  &  $-$2.25 &   15.37  \\
NGC 7089 (M2)   &14.89 $\pm$ 0.09 &   13.29 $\pm$  0.11 &    0.06 &  0.91 & $-$9.02  &  $-$1.62 &   15.49  \\
\enddata
\end{deluxetable}
\clearpage

\topmargin 3.5cm
\textheight 21cm
\begin{deluxetable}{lcccccccccccccccccc}
\rotate
\tablecaption{The selected galaxy sample}
\tablewidth{770pt}
\tabletypesize{\small}
\setlength{\tabcolsep}{0.05in}
\tablenum{8}
\tablehead{
\colhead{Galaxy}             & \colhead{T\tablenotemark{a}}             & \colhead{$A_B$\tablenotemark{a}} &
\colhead{$(V-I)_0$\tablenotemark{a}}              &
\colhead{$\bar{m}_{I}$\tablenotemark{a}}          & \colhead{$\bar{m}_{V}$}  & \colhead{Ref.}   &
\colhead{$\bar{m}_{R}$}      & \colhead{Ref.}            & \colhead{$\bar{m}_{K'}$} &
\colhead{Ref.}                & \colhead{$\bar{m}_{Ks}$} & \colhead{Ref.}  &
\colhead{$\bar{m}_{F814W}$}  &   \colhead{Ref.}          &
\colhead{$\bar{m}_{F160W}$}  & \colhead{Ref.}            & \colhead{DM}       }
\startdata
\multicolumn {19} {l}  {Cepheids distance determination} \\
\tableline
N0224 &3&0.35&1.231$\pm$0.007&23.03$\pm$0.05&25.29$\pm$0.08&1&\nodata& &18.80$\pm$0.21 &5&18.82$\pm$0.09&8&23.54$\pm$0.04&11&19.78$\pm$0.04&12& 24.44$\pm$0.10 \\
N3031 &2&0.35&1.187$\pm$0.011&26.38$\pm$0.25&29.07$\pm$0.27&1&\nodata& &\nodata       & &\nodata      & &\nodata      &  &22.99$\pm$0.05&12& 27.80$\pm$0.08 \\
N3351 & &0.12&1.225$\pm$0.014&\nodata       &\nodata       & &\nodata& &\nodata       & &\nodata      & &\nodata      &  &25.19$\pm$0.07&12& 30.01$\pm$0.08 \\
N3368 &2&0.11&1.145$\pm$0.015&28.32$\pm$0.20&\nodata       & &\nodata& &\nodata       & &\nodata      & &\nodata      &  &24.88$\pm$0.09&12& 30.20$\pm$0.10 \\
N4258 &4&0.07&1.134$\pm$0.023&27.50$\pm$0.08&\nodata       & &\nodata& &\nodata       & &\nodata      & &\nodata      &  &\nodata       &  & 29.49$\pm$0.07\tablenotemark{b}    \\
N4548 &3&0.16&1.148$\pm$0.019&29.67$\pm$0.53&\nodata       & &\nodata& &\nodata       & &\nodata      & &\nodata      &  &\nodata       &  & 31.04$\pm$0.08 \\
N4725 &2&0.05&1.209$\pm$0.023&29.14$\pm$0.32&\nodata       & &\nodata& &\nodata       & &\nodata      & &\nodata      &  &25.72$\pm$0.10&12& 30.57$\pm$0.08 \\
N7331 &3&0.39&1.120$\pm$0.017&28.72$\pm$0.14&\nodata       & &\nodata& &\nodata       & &\nodata      & &\nodata      &  &25.43$\pm$0.08&12& 30.89$\pm$0.10 \\
\tableline
\multicolumn {19} {l} {Other distance determination}  \\
\tableline
N0221 &-6&0.35&1.133$\pm$0.007&22.73$\pm$0.05&25.14$\pm$0.10&2&24.36$\pm$0.10&2&18.56$\pm$0.09&6&18.56$\pm$0.08&8&22.84$\pm$0.03&11&19.11$\pm$0.05&12&24.56$\pm$0.17 \\
N0891 & 3&0.28&1.142$\pm$0.017&27.83$\pm$0.11&\nodata       & &\nodata       & &\nodata      & &\nodata       & &\nodata      &  &\nodata       &  &29.95$\pm$0.05 \\
N1023 &-3&0.26&1.193$\pm$0.017&28.74$\pm$0.13&\nodata       & &\nodata       & &\nodata      & &\nodata       & &29.11$\pm$0.04&11&\nodata       &  &30.23$\pm$0.08 \\
N1316 &-2&0.09&1.132$\pm$0.016&29.83$\pm$0.15&32.25$\pm$0.16&3&\nodata       & &\nodata      & &\nodata       & &\nodata      &  &26.11$\pm$0.09&12&31.28$\pm$0.09 \\
N1399 &-5&0.06&1.227$\pm$0.016&30.11$\pm$0.13&32.47$\pm$0.12&3&\nodata       & &25.98$\pm$0.16&7&26.14$\pm$0.12&10&\nodata    &  &26.79$\pm$0.04&12&31.31$\pm$0.09 \\
N1404 &-5&0.05&1.224$\pm$0.016&30.20$\pm$0.16&32.48$\pm$0.12&3&\nodata       & &25.82$\pm$0.14&7&25.89$\pm$0.05&10&\nodata    &  &26.69$\pm$0.08&12&31.33$\pm$0.10 \\
N3115 &-3&0.20&1.183$\pm$0.010&28.34$\pm$0.06&\nodata       & &\nodata       & &\nodata       & &\nodata       &  &\nodata    &  &\nodata       & &30.05$\pm$0.10 \\
N3377 &-5&0.15&1.114$\pm$0.009&28.35$\pm$0.06&30.57$\pm$0.11&1&\nodata       & &\nodata       & &\nodata       &  &28.59$\pm$0.05&11&\nodata    & &30.19$\pm$0.08  \\
N3379 &-5&0.10&1.193$\pm$0.015&28.57$\pm$0.07&31.21$\pm$0.06&2&30.36$\pm$0.05&2&\nodata    & &24.72$\pm$0.06 &9 &28.81$\pm$0.11&13&25.26$\pm$0.08&12&30.23$\pm$0.06\\
N3384 &-3&0.11&1.151$\pm$0.018&28.59$\pm$0.10&\nodata       & &\nodata       & &\nodata    & &\nodata       &  &28.87$\pm$0.02&11&25.34$\pm$0.17&12&30.28$\pm$0.08 \\
N4278 &-5&0.12&1.161$\pm$0.012&29.34$\pm$0.18&\nodata       & & \nodata      & &\nodata    & &\nodata       &  &\nodata       &  &26.38$\pm$0.09&12&30.51$\pm$0.14 \\
N4374 &-5&0.17&1.191$\pm$0.008&29.77$\pm$0.09&32.02$\pm$0.09&2&31.22$\pm$0.07&2&\nodata    & &25.43$\pm$0.22&8 &30.02$\pm$0.01&14&\nodata       &  &31.20$\pm$0.10 \\
N4382 &-1&0.13&1.150$\pm$0.022&29.59$\pm$0.09&\nodata       & &\nodata       & &\nodata    & &\nodata       &  &\nodata       &  &\nodata       &  &31.07$\pm$0.11 \\
N4406 &-5&0.13&1.167$\pm$0.008&29.51$\pm$0.12&32.11$\pm$0.10&2&31.23$\pm$0.06&2&25.45$\pm$0.10&7&\nodata    &  &30.10$\pm$0.03&14&26.26$\pm$0.06&12&31.22$\pm$0.07 \\
N4472 &-5&0.10&1.218$\pm$0.011&29.62$\pm$0.07&32.26$\pm$0.06 &2&31.27$\pm$0.06&2&25.30$\pm$0.11&7&25.25$\pm$0.13&8&29.99$\pm$0.02&14&26.26$\pm$0.04&12&31.07$\pm$0.08 \\
N4486 &-4&0.10&1.244$\pm$0.012&29.72$\pm$0.14&\nodata     & &\nodata       & &\nodata       & &\nodata     & &30.07$\pm$0.03   &14&\nodata    & & 30.97$\pm$0.10 \\
N4494 &-5&0.09&1.139$\pm$0.010&29.38$\pm$0.08&\nodata     & &\nodata       & &\nodata       & &\nodata     & &\nodata         &  &\nodata    & & 30.81$\pm$0.08 \\
N4565 & 3&0.07&1.128$\pm$0.027&29.37$\pm$0.11&\nodata     & &\nodata       & &\nodata       & &\nodata     & &\nodata         &  &25.53$\pm$0.08&12&30.63$\pm$0.12\\
N4649 &-5&0.12&1.232$\pm$0.023&29.76$\pm$0.09&\nodata     & &\nodata       & &\nodata       & &\nodata     & &30.21$\pm$0.07  &14&\nodata    & & 31.09$\pm$0.08 \\
N4660 &-5&0.14&1.154$\pm$0.015&28.82$\pm$0.17&\nodata     & &\nodata       & &\nodata       & &\nodata     & &30.11$\pm$0.09  &14&\nodata    & & 30.70$\pm$0.08 \\
N5102 &-3&0.24&0.976$\pm$0.016&25.49$\pm$0.09&\nodata     & &\nodata       & &\nodata       & &\nodata     & &\nodata         &  &\nodata    & & 27.89$\pm$0.13 \\
N5128 &-2&0.50&1.078$\pm$0.016&26.05$\pm$0.11&28.12$\pm$0.06&4&\nodata     & &\nodata       & &\nodata     & &\nodata         &  &\nodata    & & 27.90$\pm$0.08 \\
N5195 & 0&0.16&1.056$\pm$0.019&27.25$\pm$0.25&\nodata       & &\nodata     & &\nodata       & &\nodata     & &\nodata         &  &\nodata    & &  29.37$\pm$0.08 \\
\enddata
\tablecomments{References:
(1)=Ajhar \& Tonry 1994;
(2)=Tonry, Ajhar \& Luppino 1990;
(3)=Blakeslee, Vazdekis \& Ajhar 2001;
(4)=Tonry \& Schetcher 1990;
(5)=Jensen et al. 1996;
(6)=Luppino \& Tonry 1993;
(7)=Jensen et al. 1998;
(8)=Pahre \& Mould 1994;
(9)=Mei et al. 2001;
(10)=Liu, Graham \& Charlot 2001;
(11)=Ajhar et al. 1997;
(12)=Jensen et al. 2003;
(13)=Pahre et al. 1999;
(14)=Neilsen \& Tsvetanov 2000}
\tablenotetext{a}{Data from T01}
\tablenotetext{b}{Freedman et al. 2001, Table~3}
\end{deluxetable}
\clearpage

\begin{deluxetable}{lrrrrrcccc}
\rotate
\tabletypesize{\small}
\tablenum{9}
\tablecaption{Age and metallicity extimations of selected galaxies}
\tablewidth{550pt}
\tablehead{
\colhead{Galaxy}   & \colhead{DM} &  
\colhead{Age\tablenotemark{a}}  & \colhead{[Fe/H]\tablenotemark{a}} &
\colhead{Age\tablenotemark{b}}  & \colhead{[Fe/H]\tablenotemark{b}} &
\colhead{Age$^{SBF}_{mag}$}  & \colhead{[Fe/H]$^{SBF}_{mag}$} &
\colhead{Age$^{SBF}_{color}$} & \colhead{[Fe/H]$^{SBF}_{color}$}\\
\colhead{(1)}  & \colhead{(2)} &  
\colhead{(3)}  & \colhead{(4)} &
\colhead{(5)}  & \colhead{(6)} &
\colhead{(7)}  & \colhead{(8)} &
\colhead{(9)}  & \colhead{(10)}}
\startdata
NGC 0221 & $24.44\pm0.10$ & 3.8      & $-$0.04     &  4.9$\pm$1.3 & $-$0.01$\pm$0.05 & 5-9       & $\sim$0.00    & 6  & 0.04 \\
NGC 0224 & $24.44\pm0.07$ & 5.1      &    0.42     &  \nodata     & \nodata          & $\sim$15  & $\sim$0.00    & 12 & 0.15 \\
NGC 0891 & $29.96\pm0.06$ & 11       & $-$0.47     &  \nodata     & \nodata          & 11-13     & $\sim-$0.27   & \nodata & \nodata   \\
NGC 1316 & $31.49\pm0.03$ & 3.4      &    0.25     &  \nodata     & \nodata          & $\sim$5   & $\sim$0.00    & \nodata & \nodata\\
NGC 1399 & $31.49\pm0.03$ & \nodata  & $>$0.5      &  \nodata     & \nodata          & $\gsim10$ & $\gsim$0.00   & 12 & 0.12\\
NGC 1404 & $31.49\pm0.03$ & 5.9      &    0.47     &  \nodata     & \nodata          & 5-9       & $\gsim$0.00   & 9 & 0.20 \\
NGC 3115 & $30.14\pm0.16$ & \nodata  & $>$0.5      &  \nodata     & \nodata          & $\sim$13  & 0.00$/-$0.27  & \nodata & \nodata\\
NGC 3377 & $30.09\pm0.04$ & 4.1      &    0.20     & 5.9$\pm$1.2  & $-$0.28$\pm$0.04 & $\sim$5   & 0.00          & \nodata & \nodata\\
NGC 3379 & $30.09\pm0.04$ & 9.3      &    0.16     & 13.2$\pm$2.4 & $-$0.17$\pm$0.04 & 9-13      & $\sim$0.00    &  12 & 0.00 \\
NGC 3384 & $30.09\pm0.04$ & \nodata  & $>$0.5      &  \nodata     & \nodata          & $\sim$5   & 0.00/0.33\tablenotemark{c}     & \nodata & \nodata\\
         &                &          &             &              &                  & 13        & -0.27\tablenotemark{d}         & \\
NGC 4278 & $31.11\pm0.09$ & 10.7     &    0.14     &  \nodata     & \nodata          & 5-9       & 0.00\tablenotemark{c}          & \nodata & \nodata\\
         &                &          &             &              &                  & 13-15     & -0.27\tablenotemark{d}         & \\
NGC 4374 & $30.97\pm0.04$ & 11.8     &    0.12     & 14.4$\pm$2.8 & $-$0.19$\pm$0.05 & 5-11      & $\gsim$0.00   & 8 & 0.15\\
NGC 4382 & $30.97\pm0.04$ & 1.6      &    0.44     &  \nodata     & \nodata          & $\lsim$5  & $\gsim$0.33   & \nodata & \nodata\\
NGC 4472 & $31.01\pm0.04$ & 8.5      &    0.24     & 8.4$\pm$2.7  &  0.00$\pm$0.06   & $\gsim$7  & $\gsim$0.00   & 9 & 0.20 \\
NGC 4649 & $31.54\pm0.08$ & 11       &    0.30     & 18.3$\pm$2.8 & $-0.22\pm0.04$   & \nodata   & \nodata       & \nodata & \nodata        \\
\enddata
\tablenotetext{a}{Terlevich \& Forbes 2002, Table~2}
\tablenotetext{b}{Trager et al. 2000, Table~6B: model 4}
\tablenotetext{c}{Estimations from $\bar{M}_{I}$}
\tablenotetext{d}{Estimations from $\bar{M}_{F160W}$}
\end{deluxetable}

\begin{deluxetable}{c|rlrlll}
\tabletypesize{\small}
\tablenum{A1}
\tablewidth{350pt}
\tablecaption{SBF predictions for a stellar population with $t=15~Gyr$ and $Z=Z_{\sun}$ as function of the 
total number of stars $N_{star}$. The last row indicates the SBF amplitudes as computed by using the 
``usual'' approach.}
\tablehead{ $N_{star}$ & $V_{tot}$ & \colhead{$\bar{M}_{B}$} & \colhead{$\bar{M}_{V}$} & \colhead{$\bar{M}_{I}$} 
& \colhead{$\bar{M}_{K}$} & \colhead{$(V-I)_0$}}
\startdata
$10^3$			& 0.455  & 2.236  &  0.710  & -1.551 & -5.373   & 1.055 \\
$10^4$			& -2.292 & 2.365  &  0.795  & -1.504 & -5.408   & 1.192 \\
$10^5$			& -4.822 & 2.353  &  0.781  & -1.513 & -5.402   & 1.233 \\
$10^6$			& -7.325 & 2.407  &  0.839  & -1.473 & -5.403   & 1.235 \\
$5\cdot10^6$	        & -9.074 & 2.395  &  0.831  & -1.483 & -5.410   & 1.235 \\
$2.6\cdot10^7$		& -10.847& 2.469  &  0.910  & -1.474 & -5.414   & 1.235 \\
$2.1\cdot10^8$		& -13.098& 2.589  &  0.989  & -1.391 & -5.365   & 1.235 \\	
\hline
``usual'' 		&\nodata & 2.599  &  1.040  & -1.371 & -5.381   & \nodata\\
\enddata
\end{deluxetable}
\clearpage

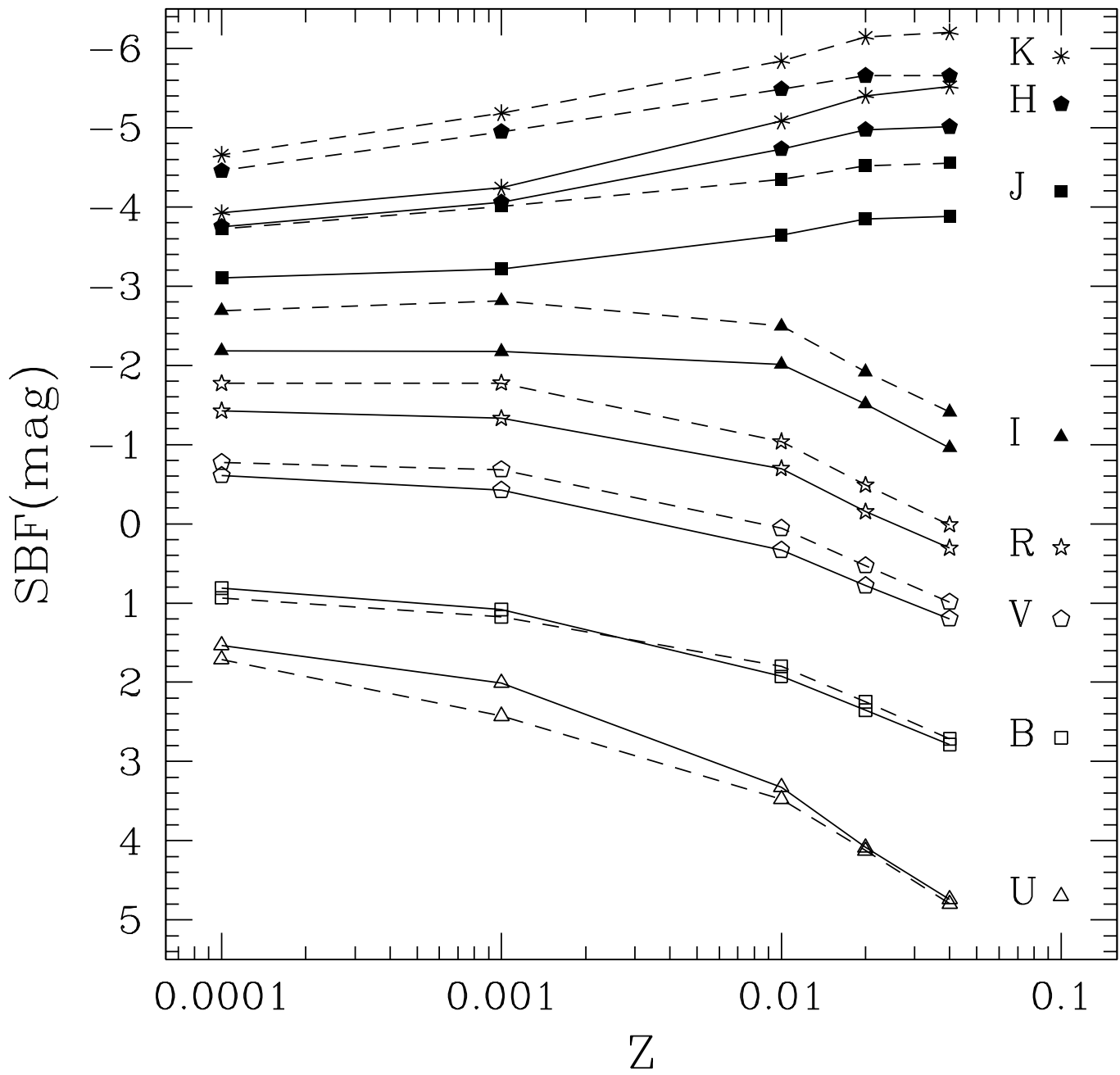
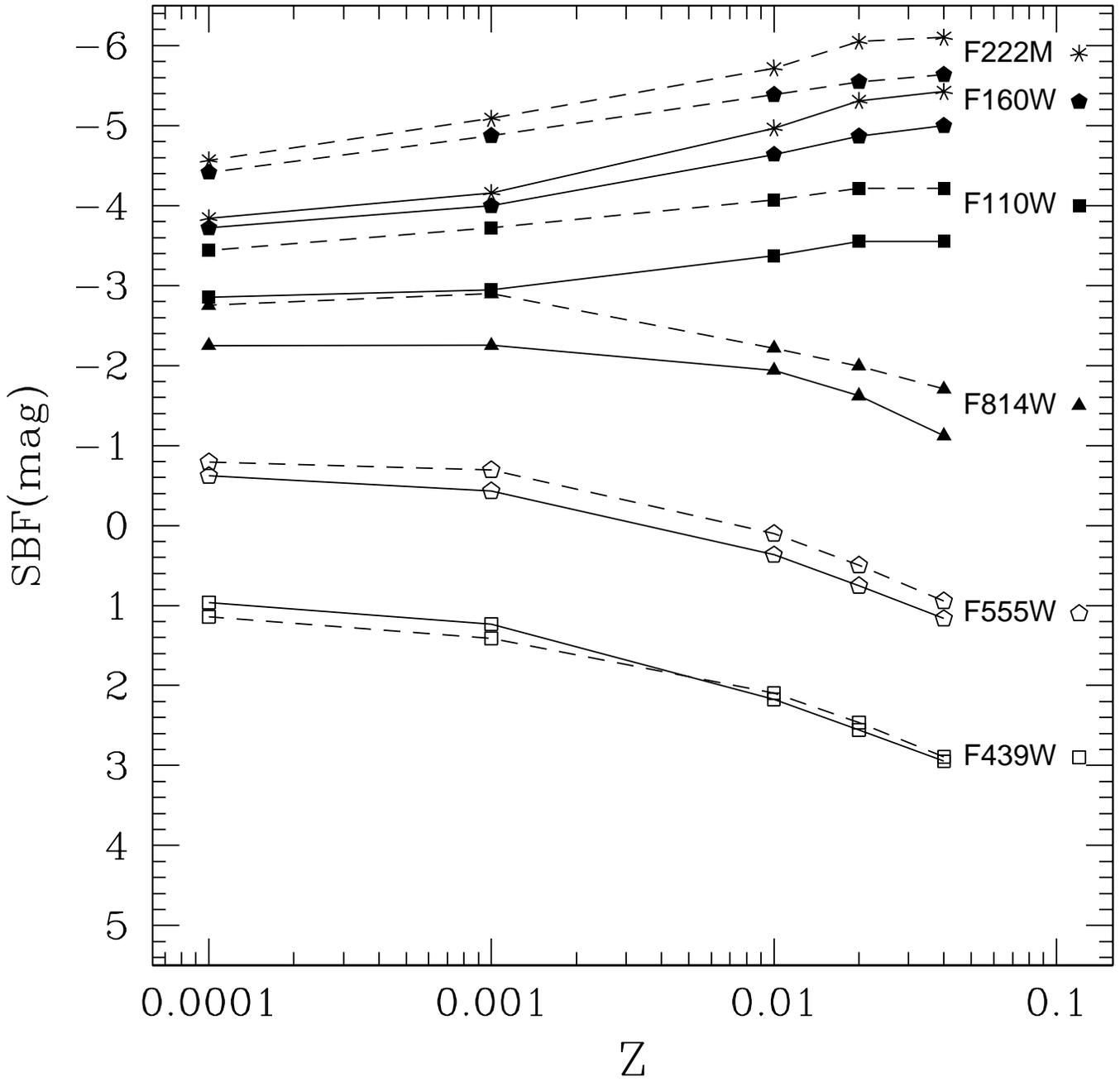
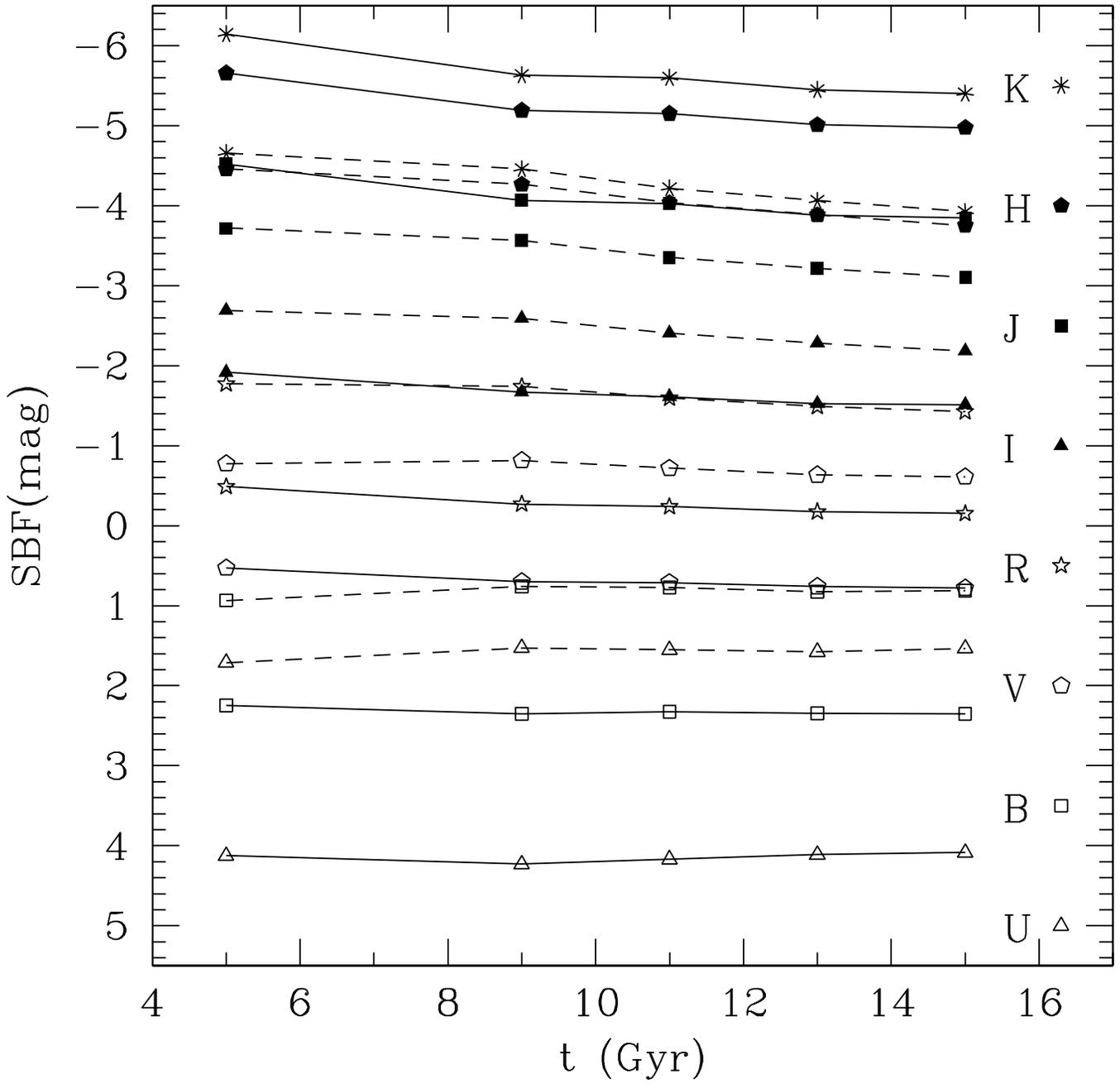
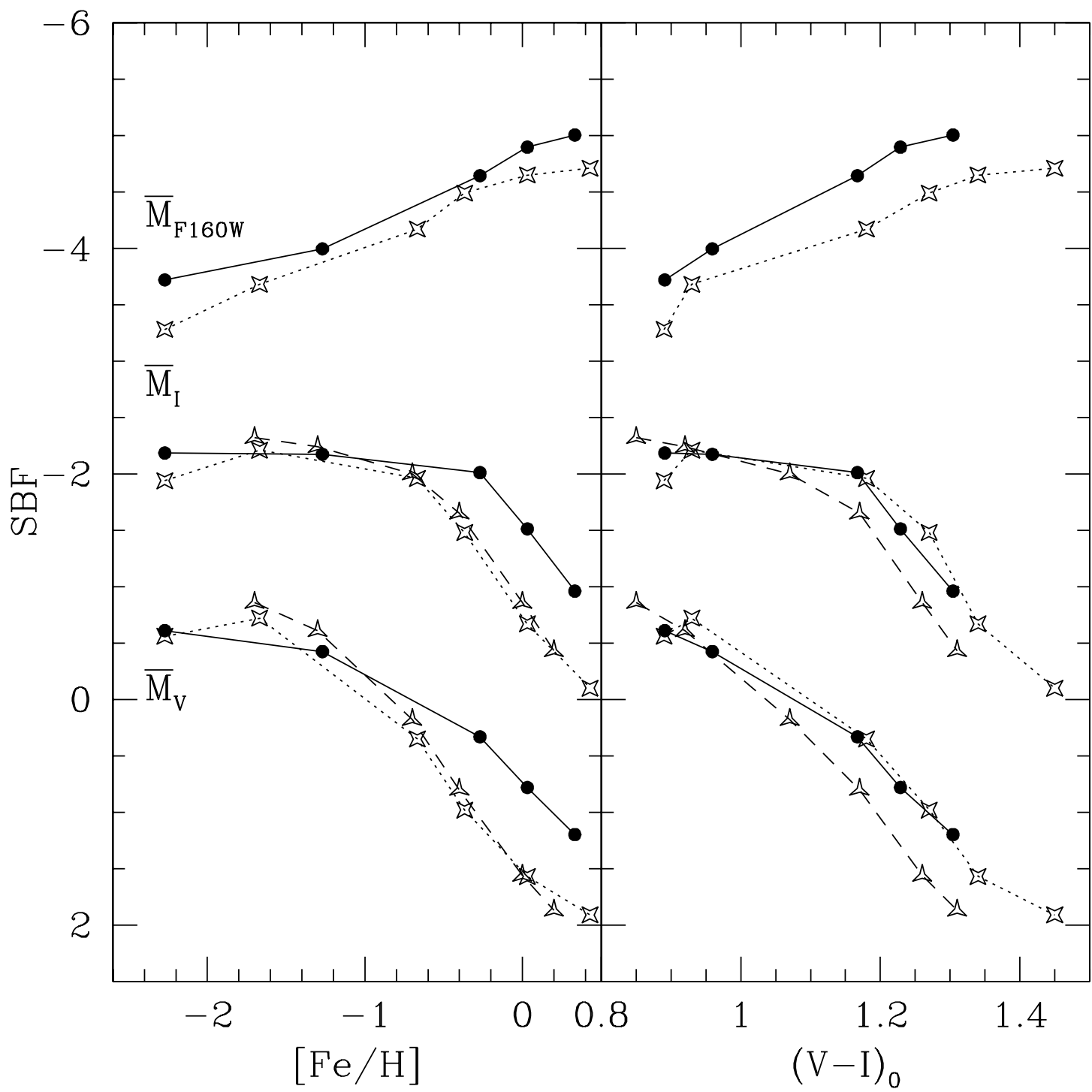
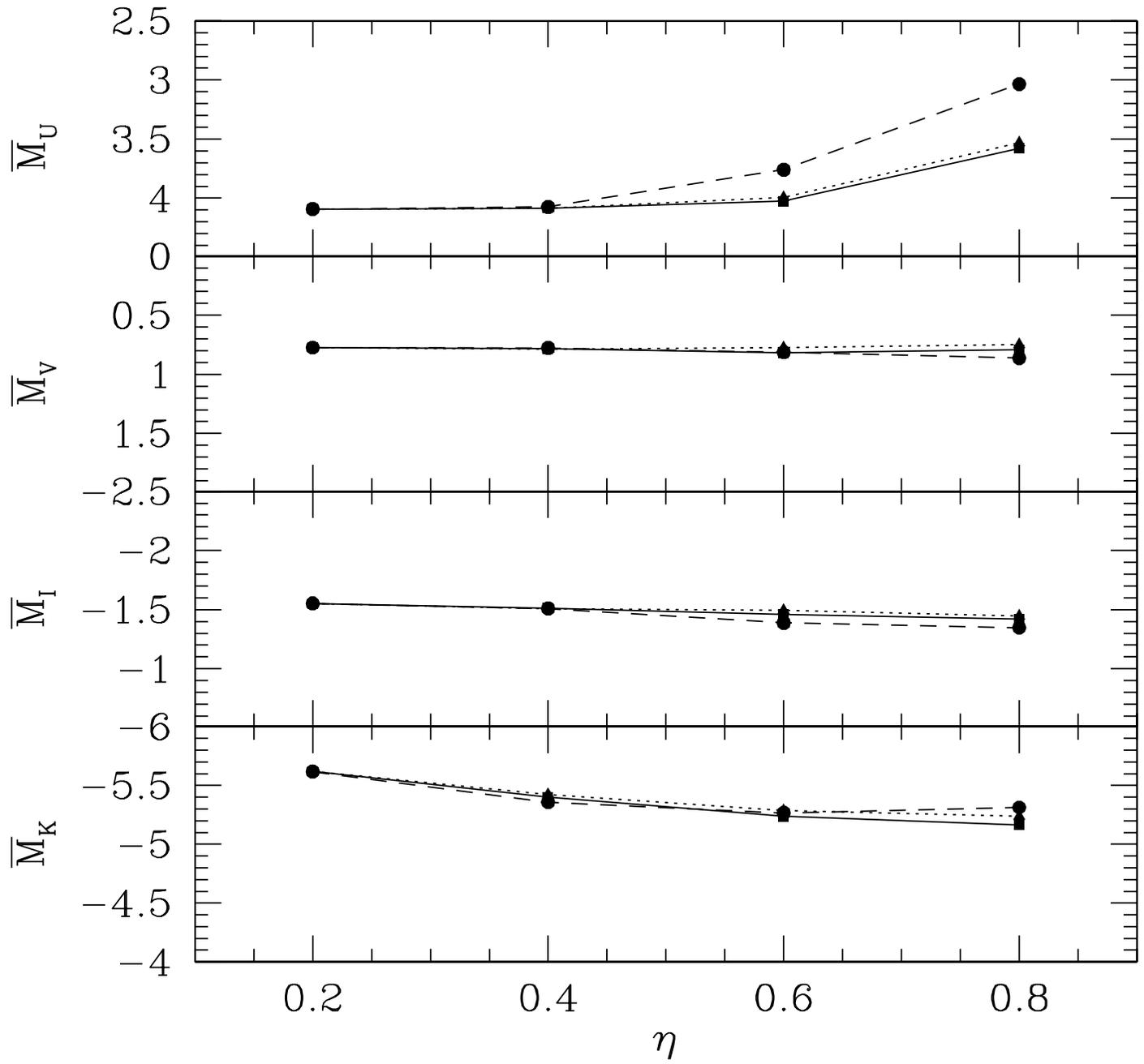
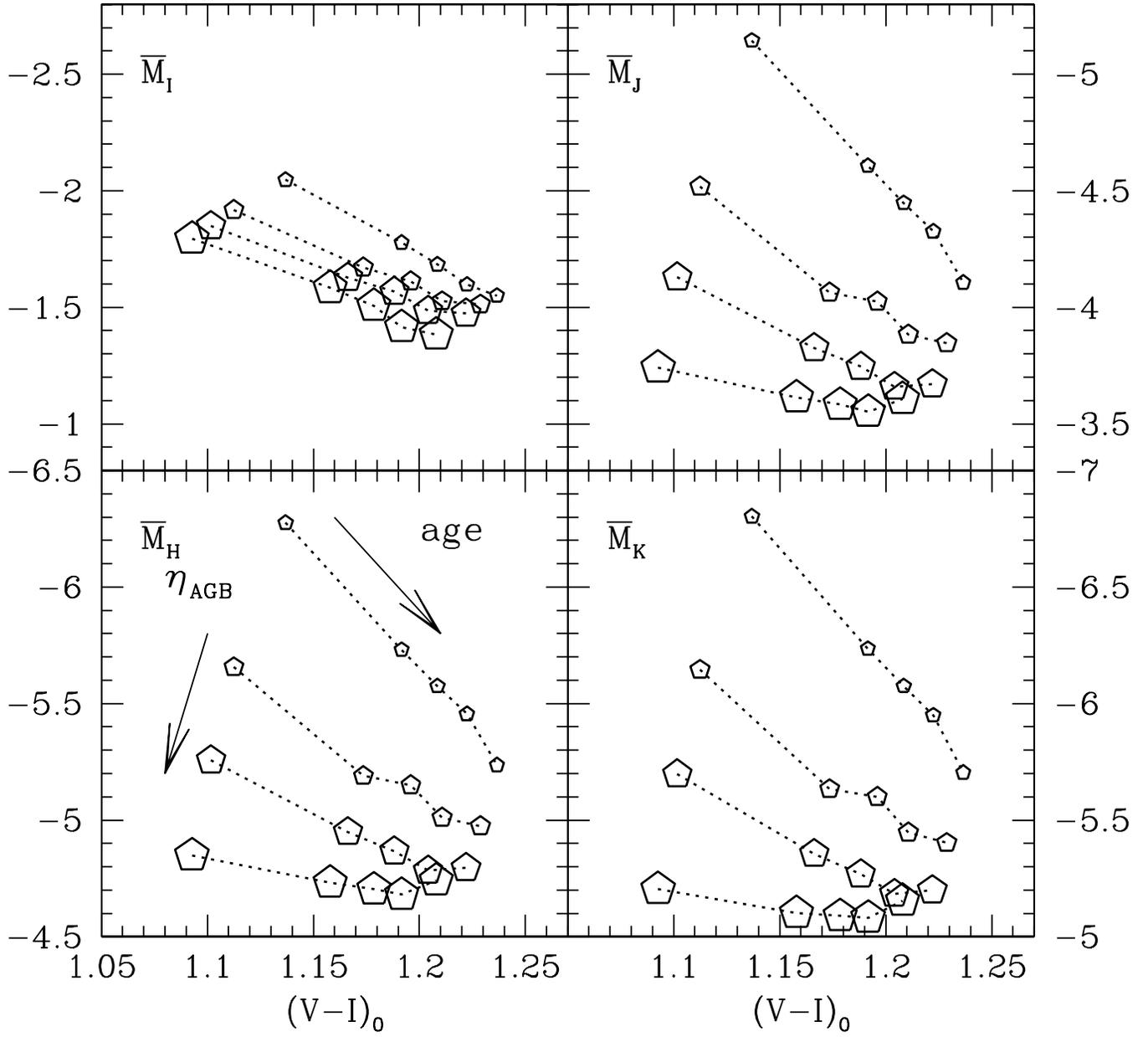
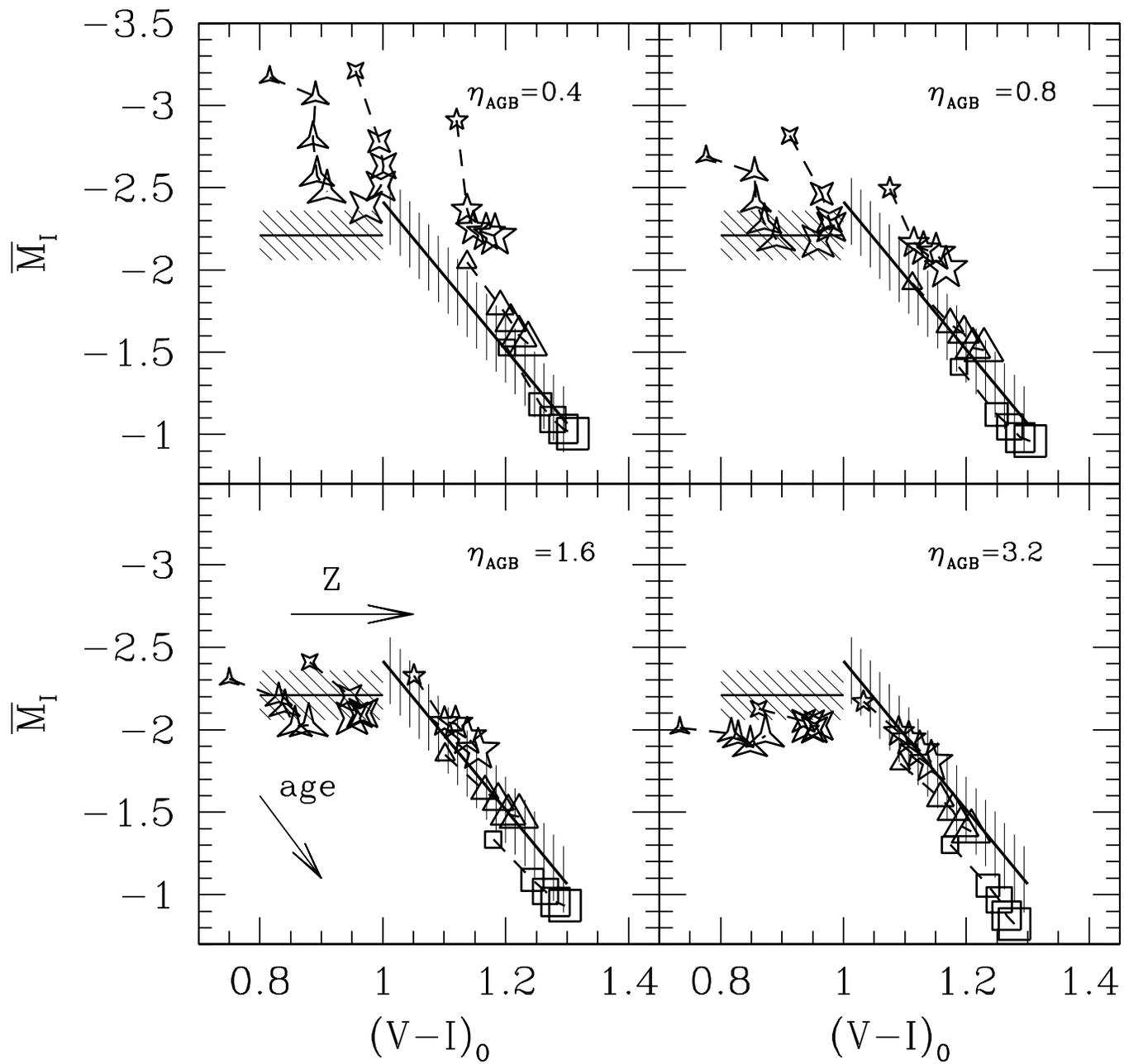
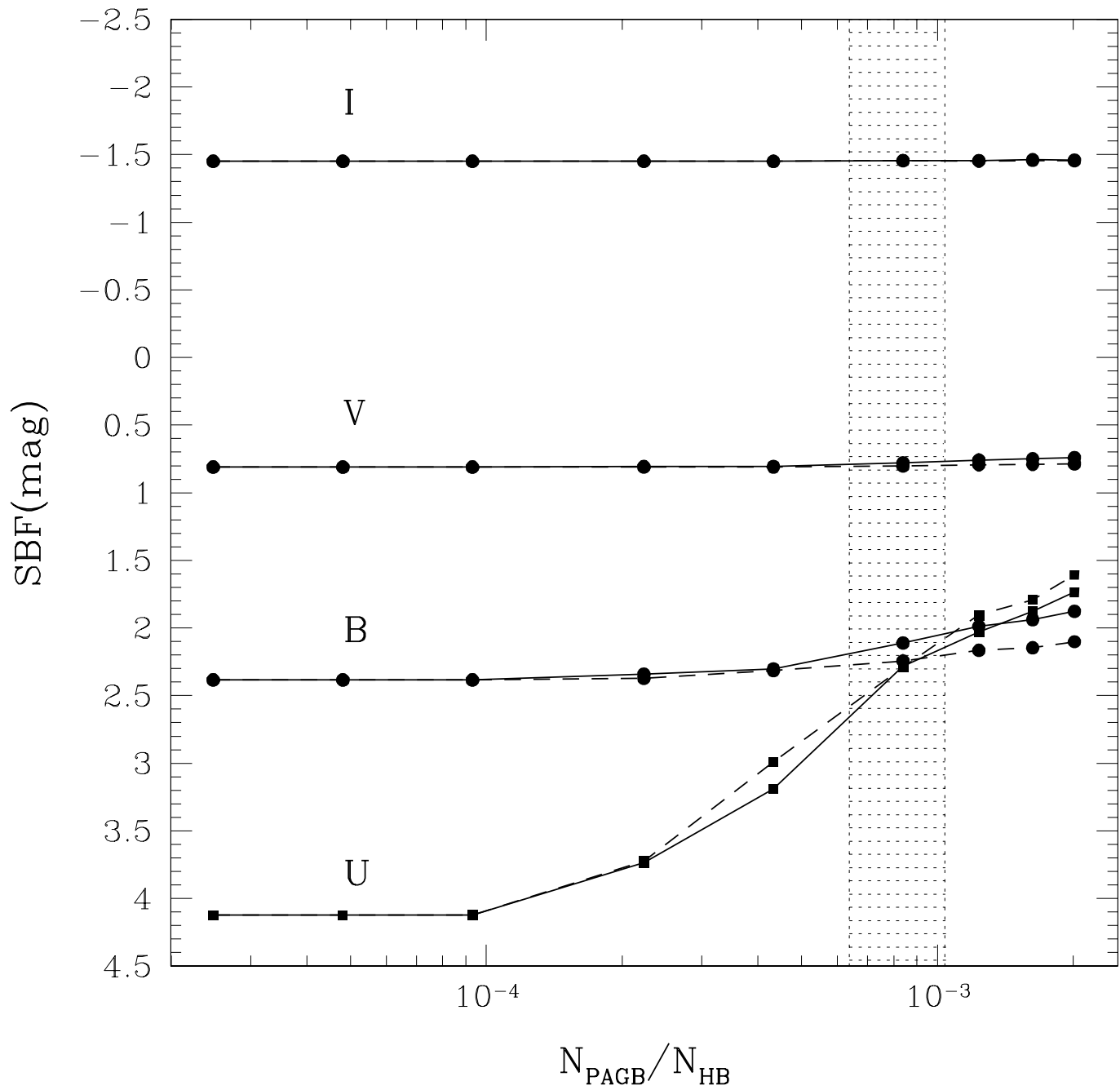
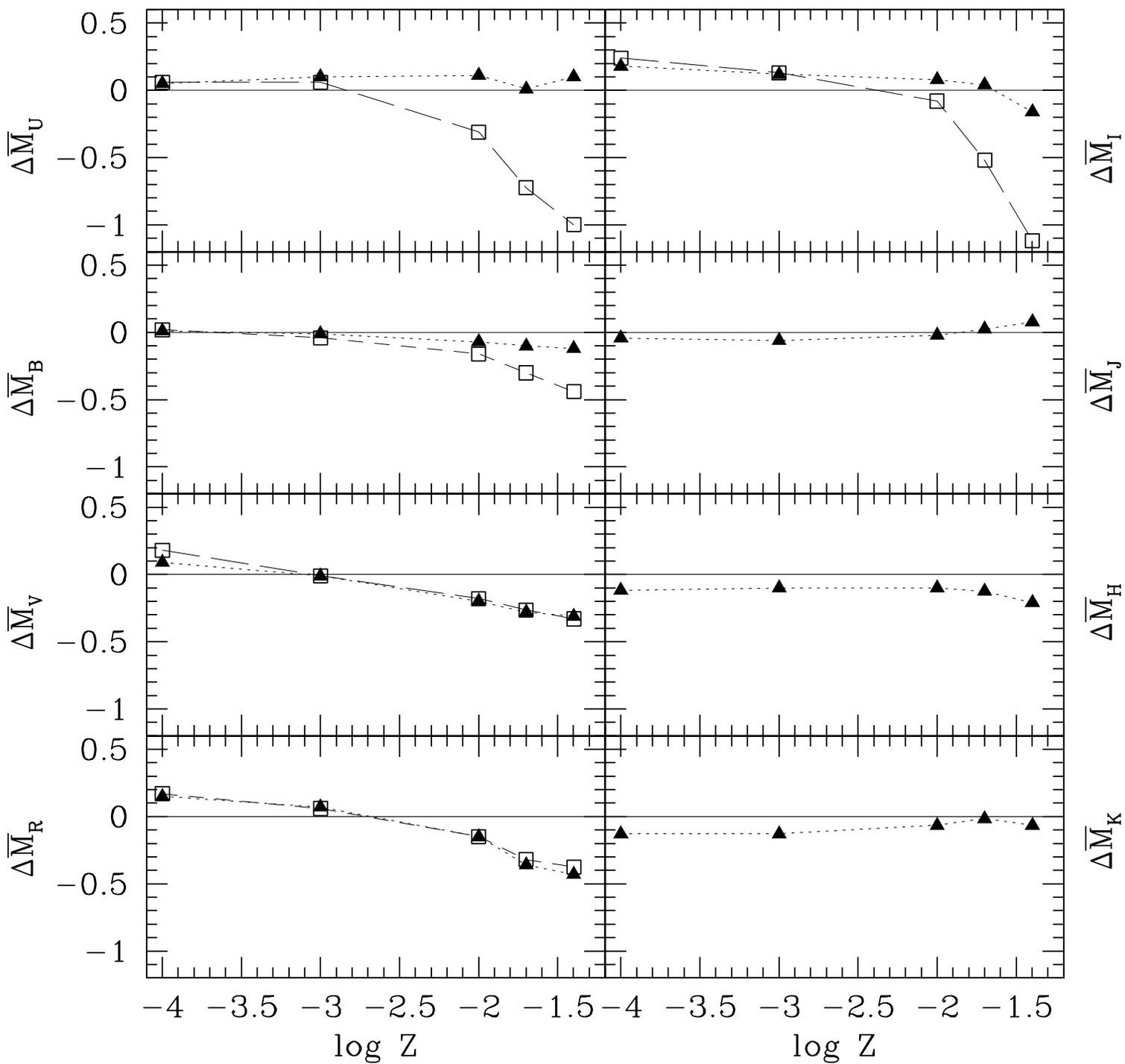
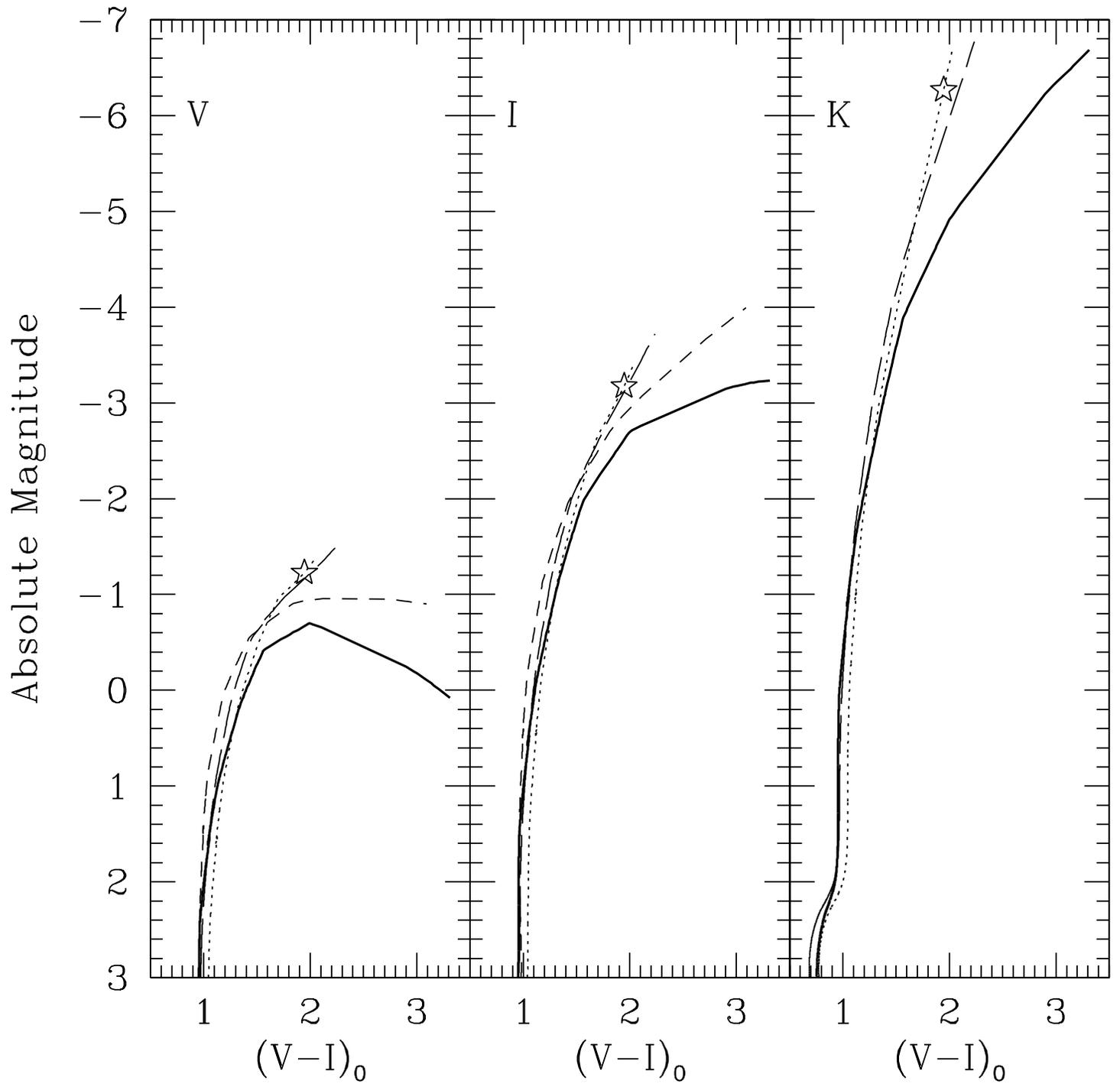
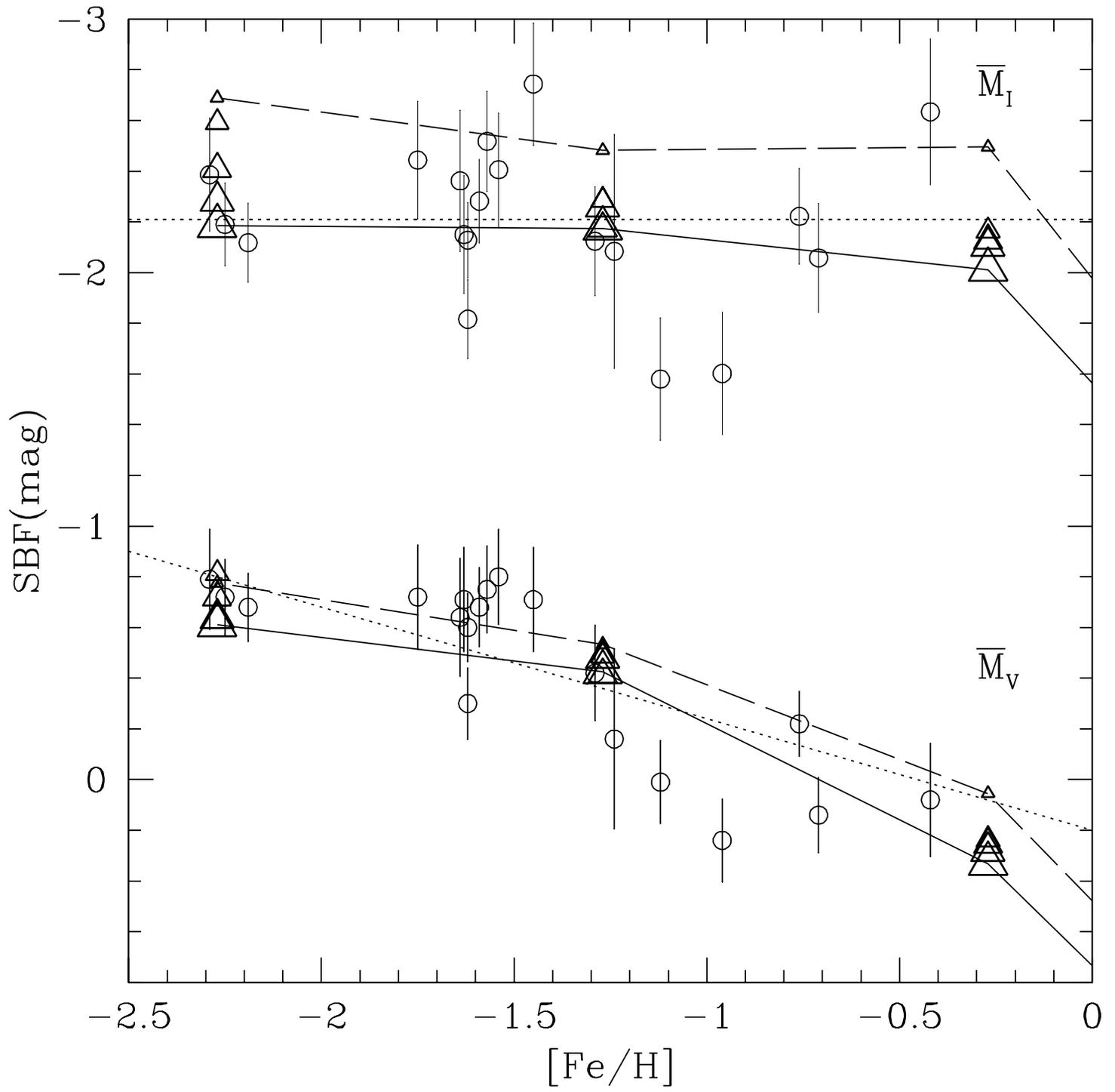
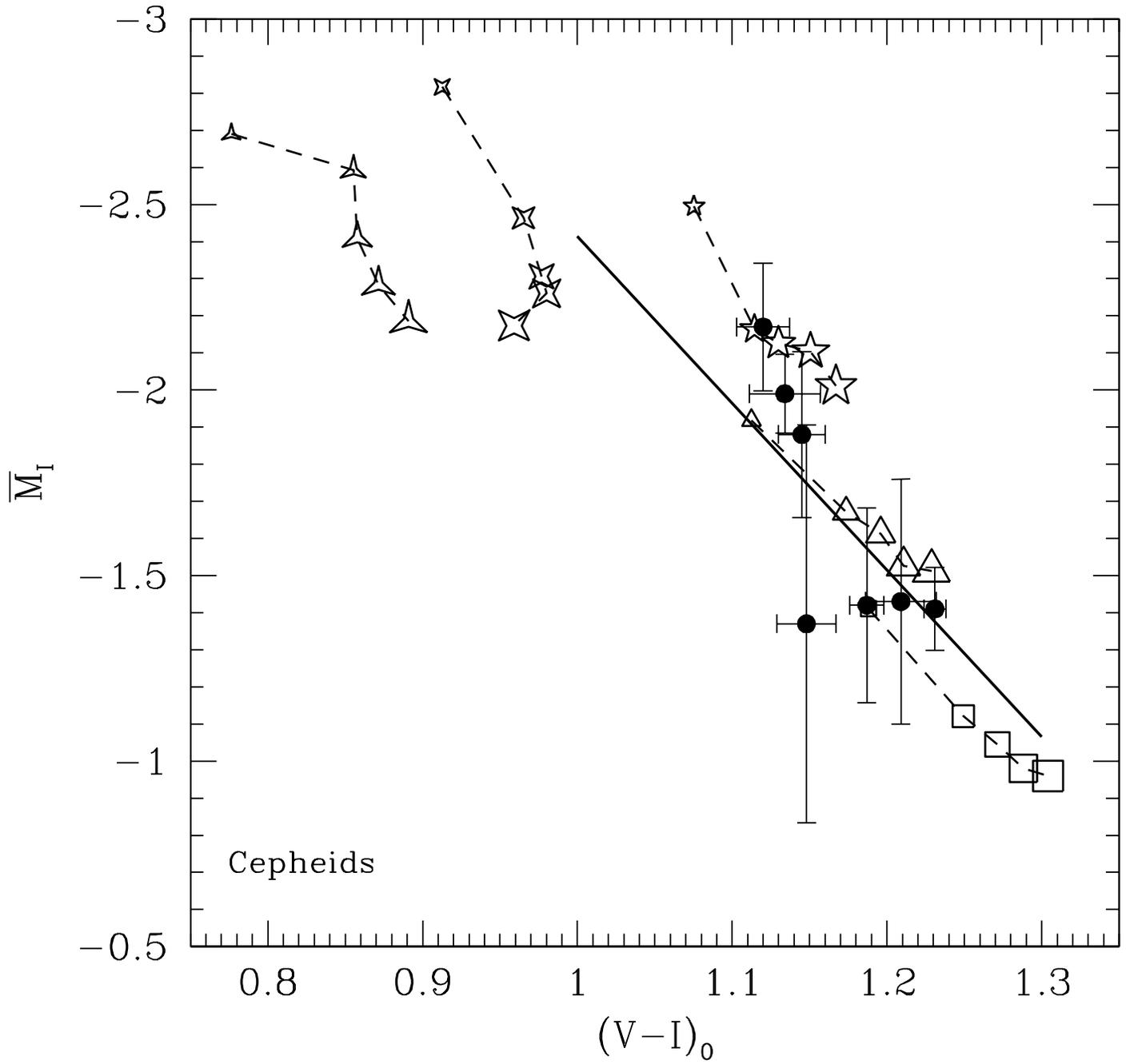
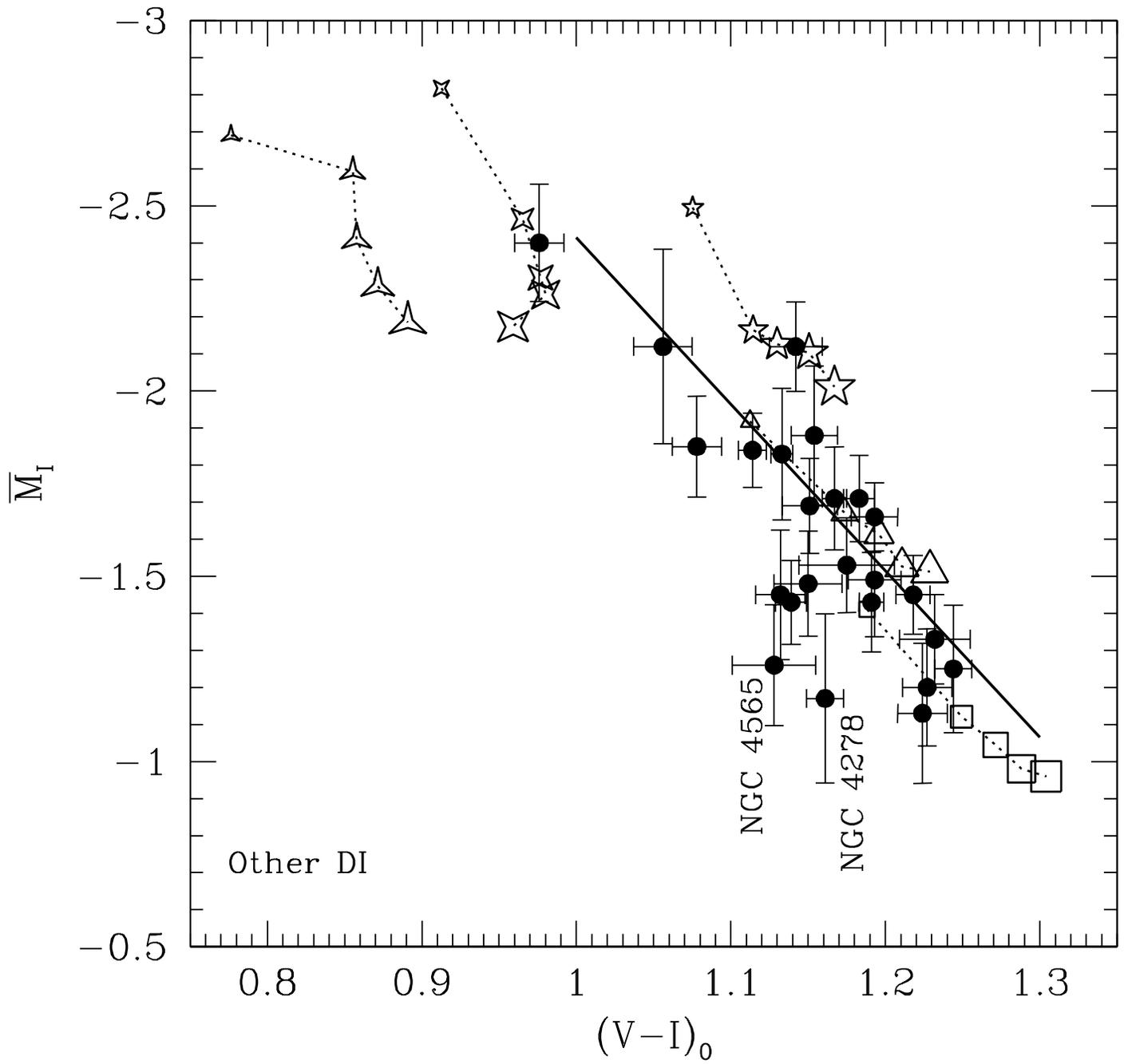
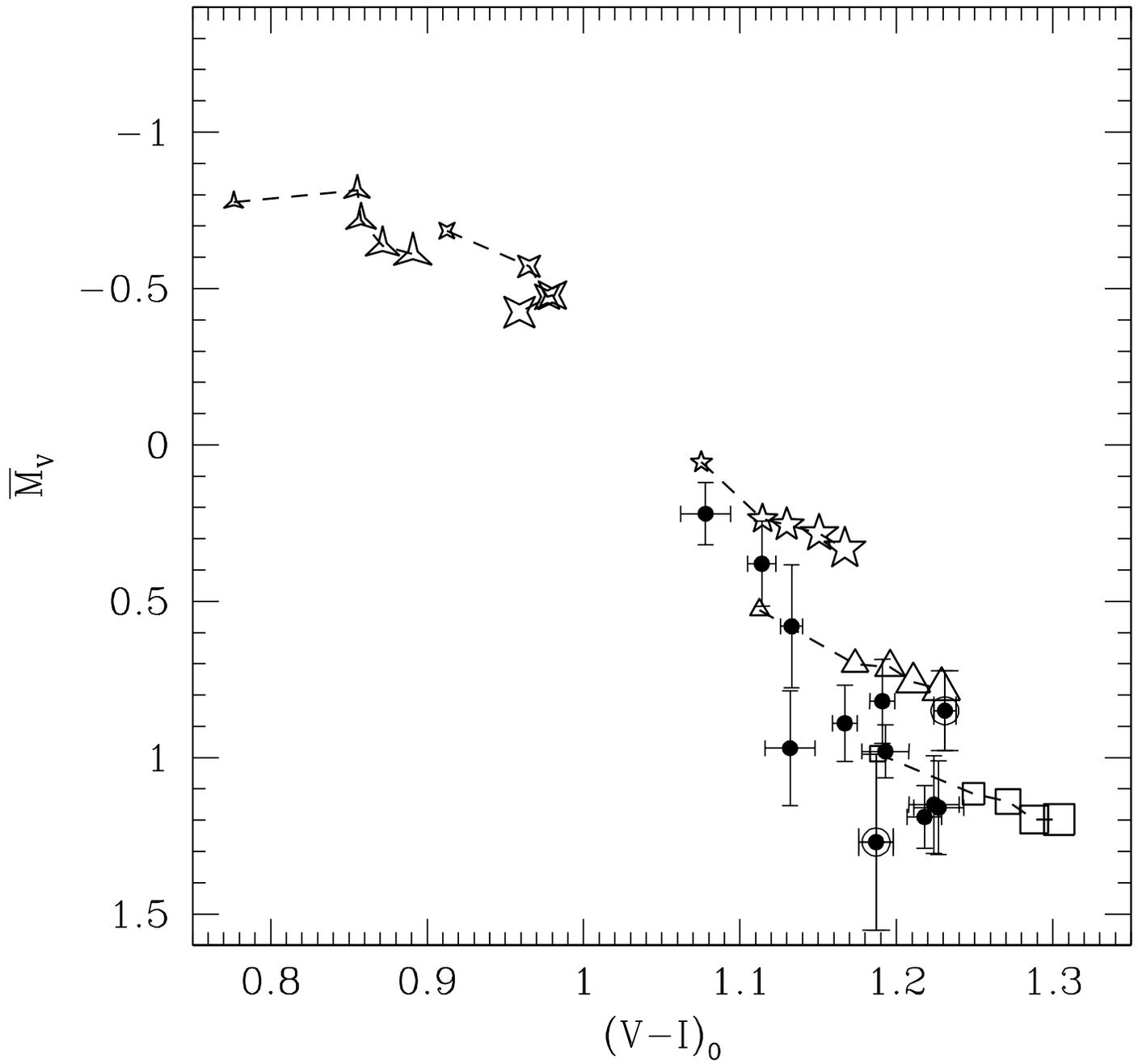
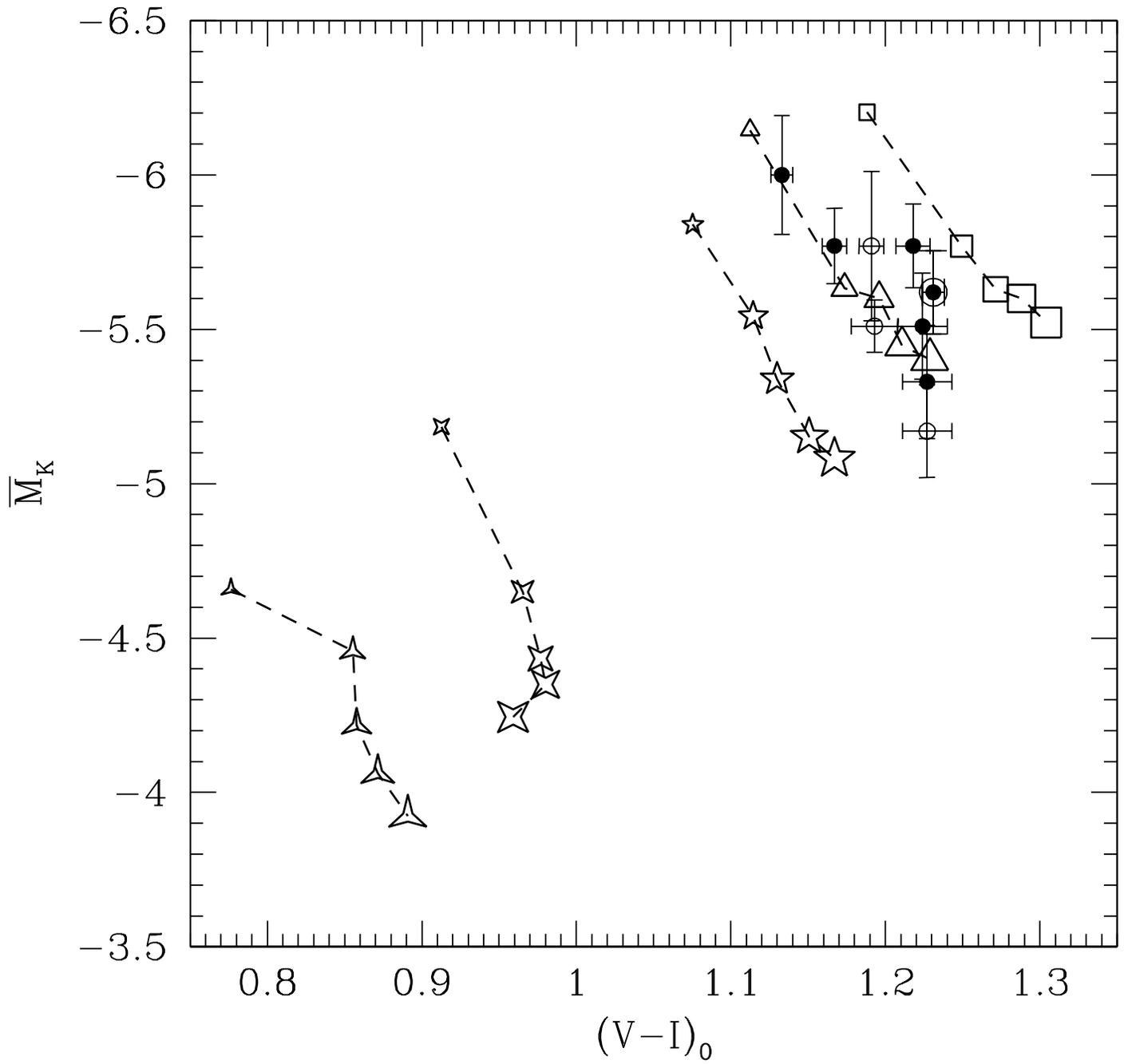
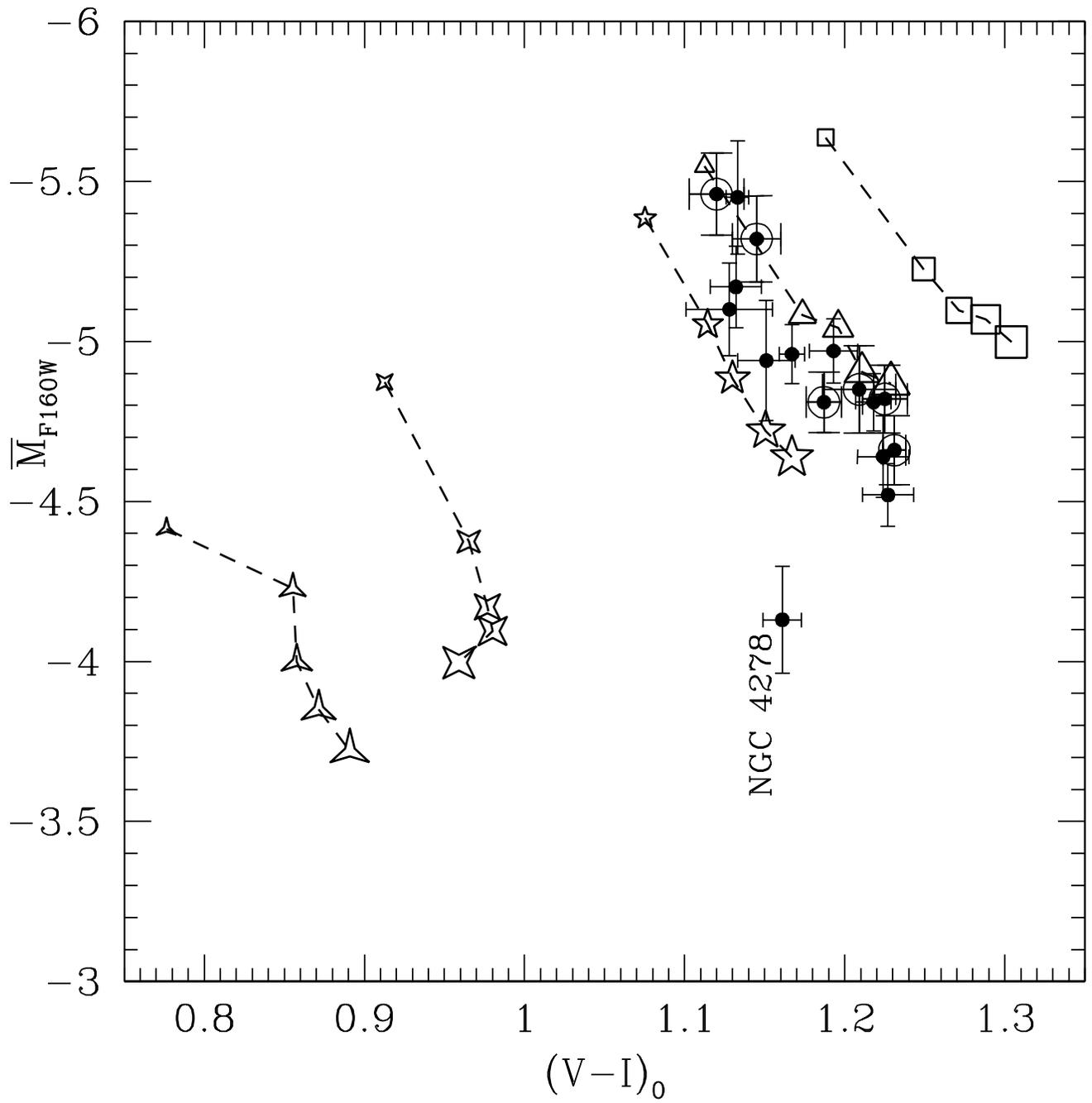
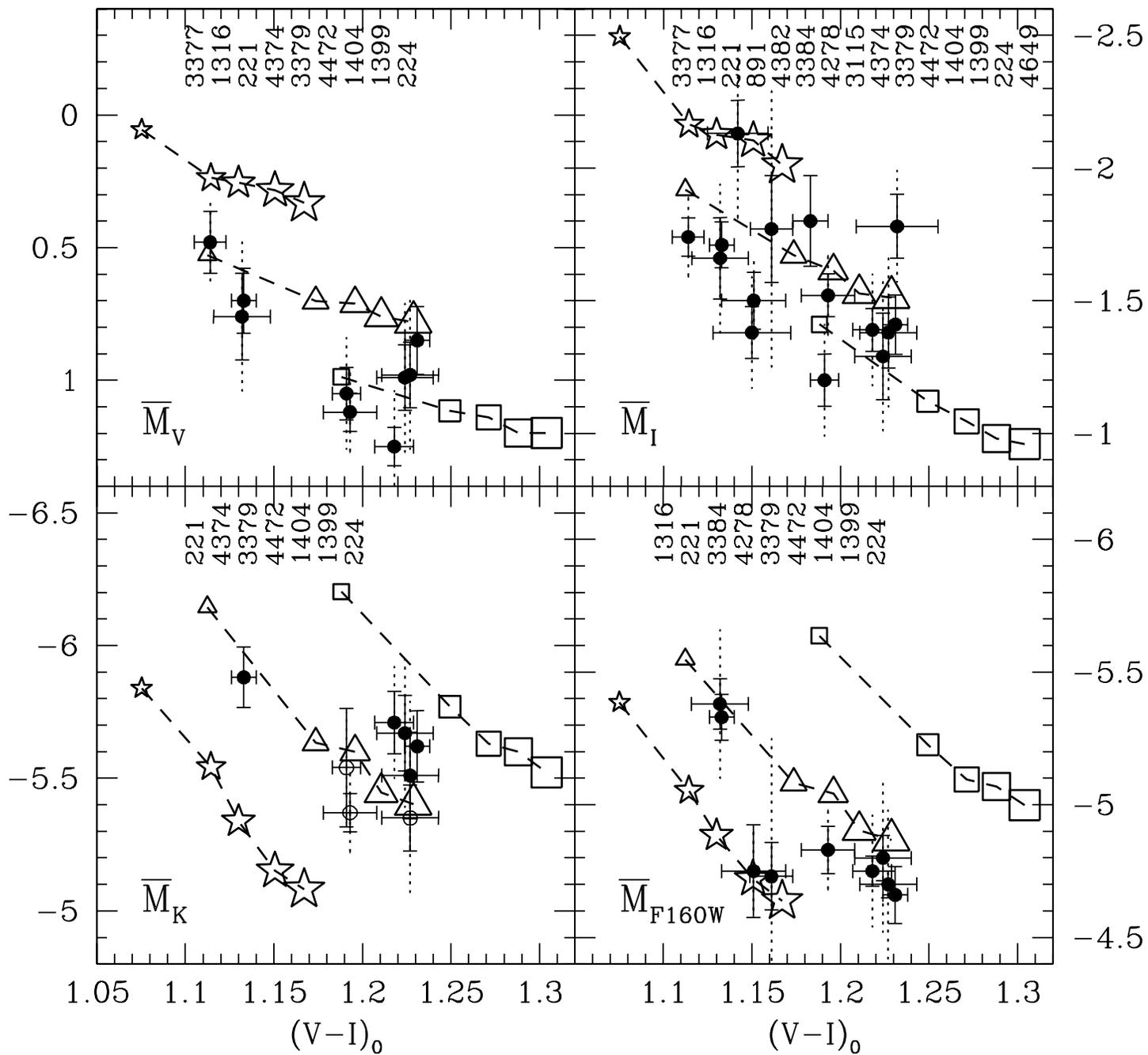
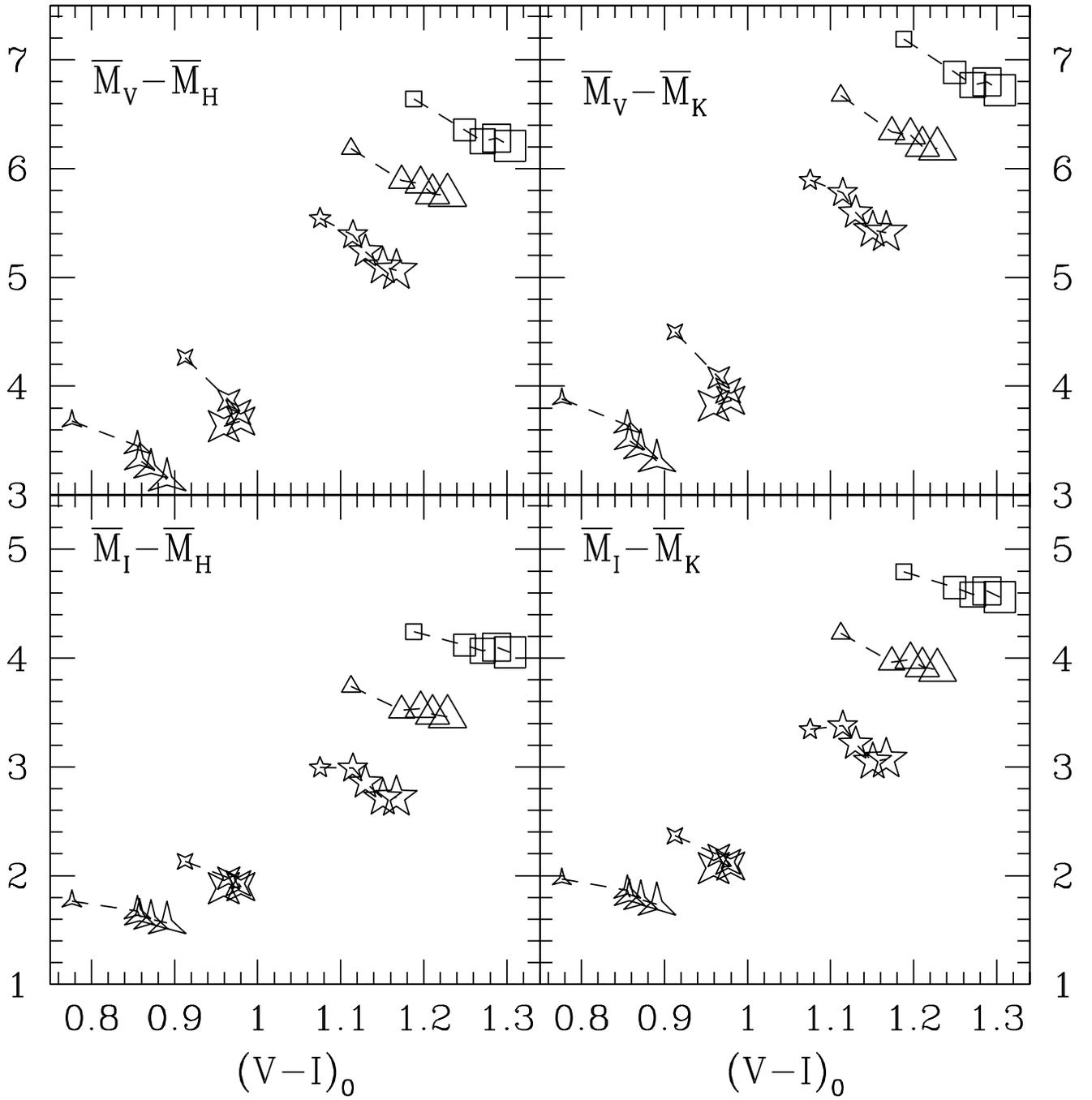
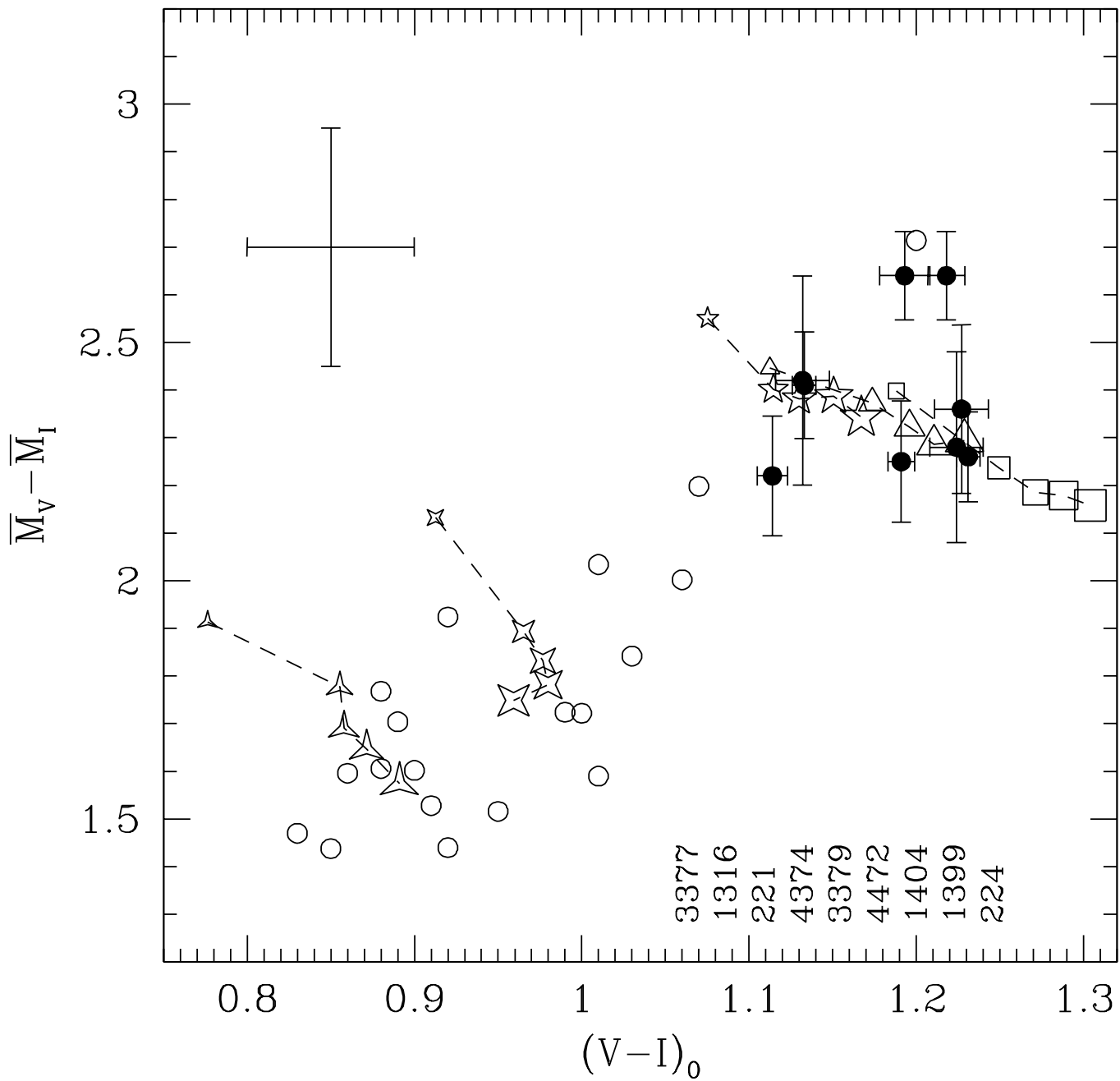
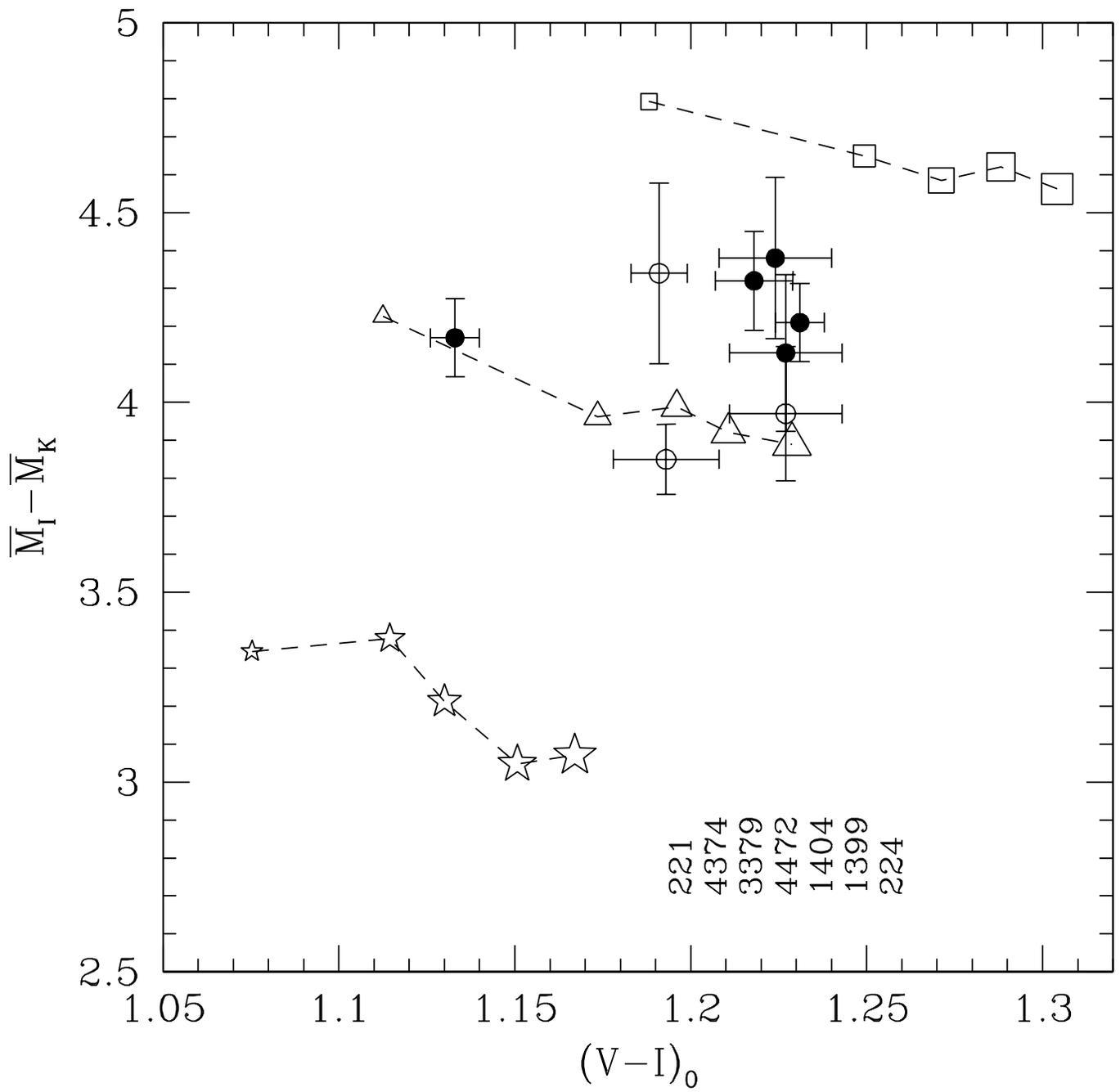
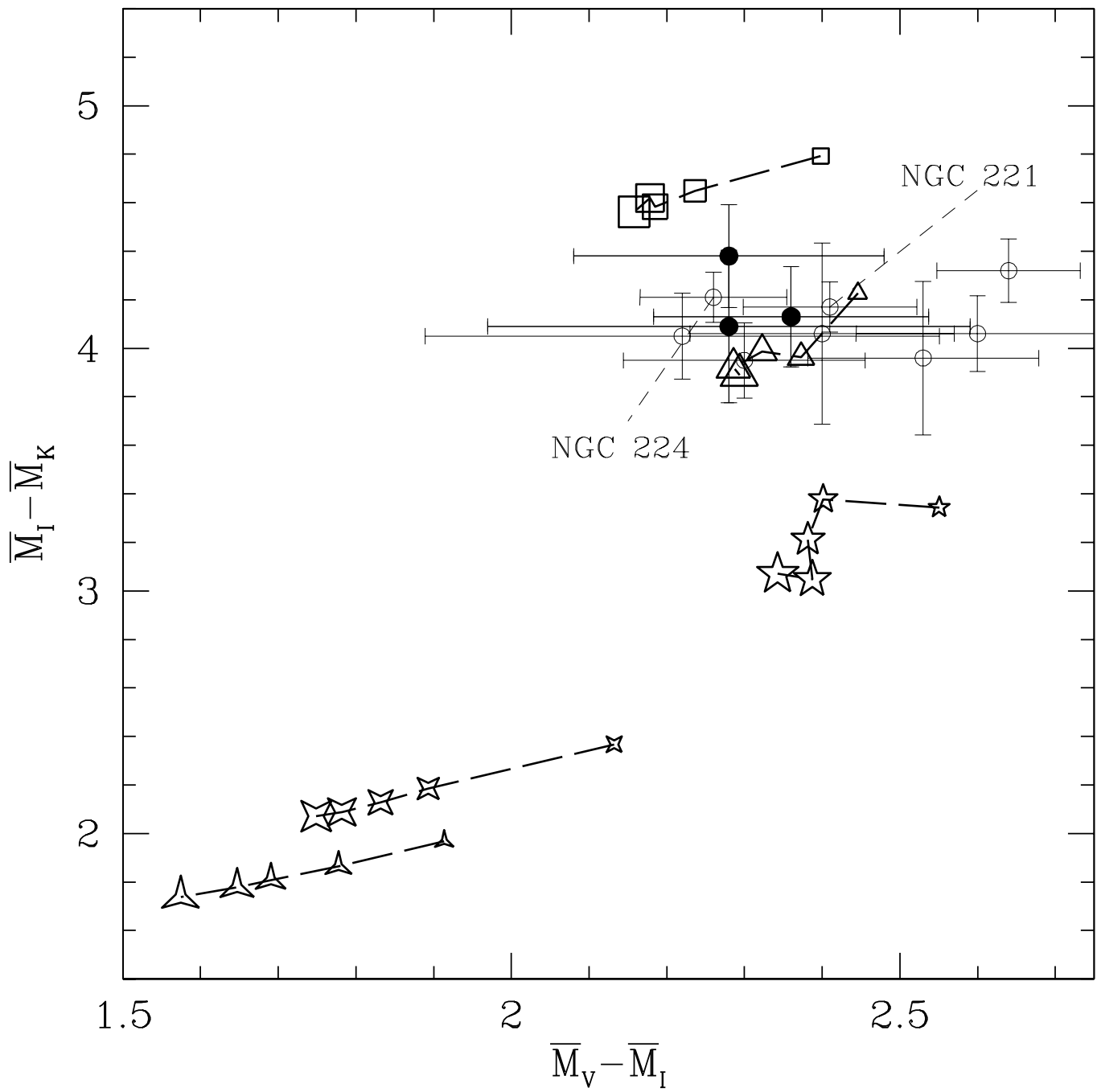
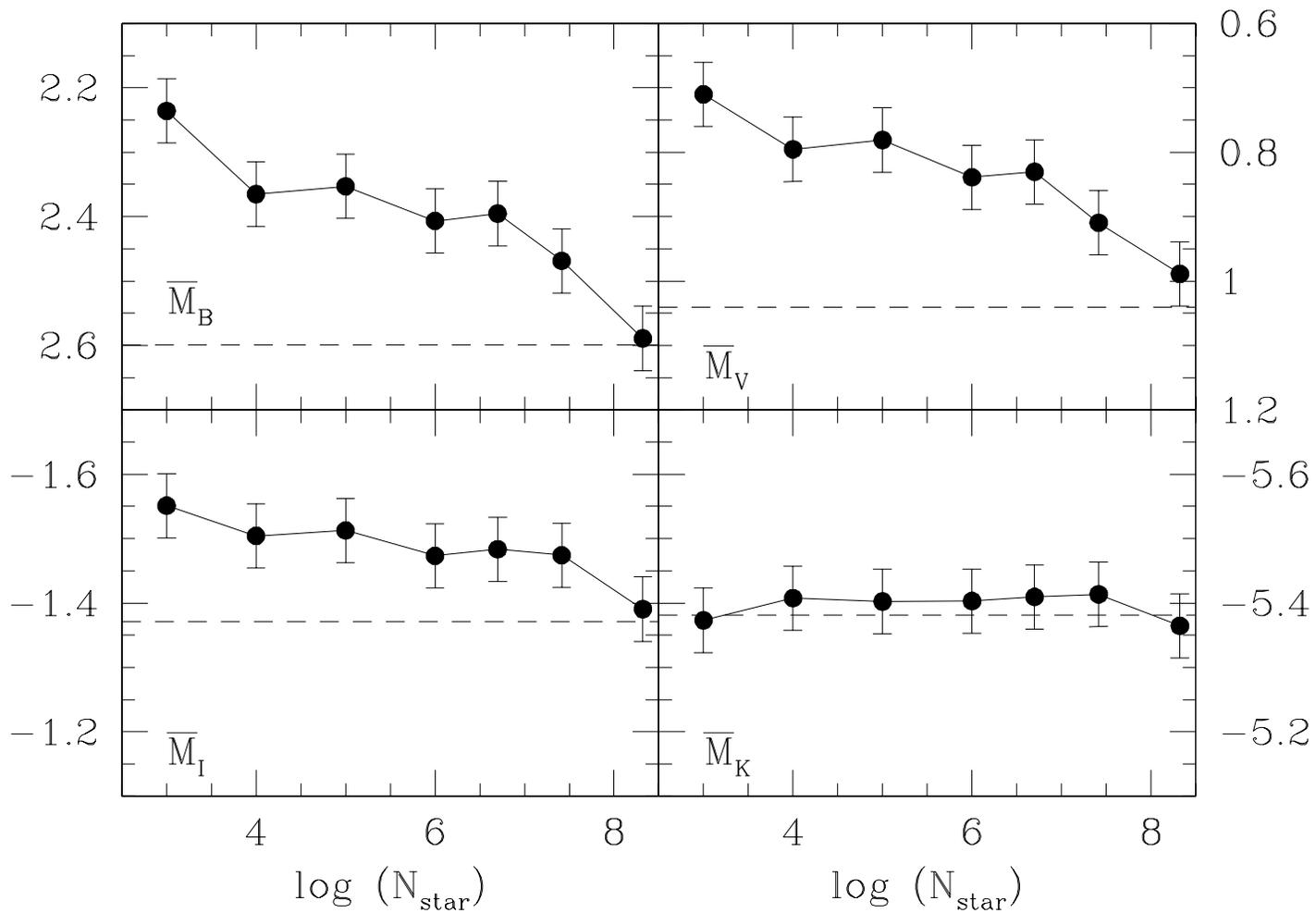
\end{document}